\documentclass[dvipsnames]{article} %

\usepackage{colm2024_conference}

\usepackage{microtype}
\usepackage{hyperref}
\usepackage{url}
\usepackage{booktabs}
\usepackage{graphicx}
\usepackage{wrapfig}
\usepackage{bbm}
\usepackage{amsmath}
\usepackage{amssymb}
\usepackage{mathtools}
\usepackage{amsthm}
\usepackage{pifont}
\usepackage{footmisc}
\usepackage{enumitem}
\usepackage{natbib}
\usepackage{xcolor}
\usepackage{multirow}
\usepackage{setspace}
\usepackage{enumitem}
\usepackage{color, colortbl}
\usepackage{mathtools}
\usepackage{cleveref}
\theoremstyle{plain}

\usepackage{hyperref}
\usepackage{listings}
\usepackage{url}
\usepackage{minted}
\usepackage{color,graphicx}
\usepackage{siunitx}
\usepackage{tabularx}
\usepackage{soul}
\usepackage{graphics} 
\usepackage{bookmark}
\usepackage[dvipsnames, svgnames, x11names]{xcolor}
\usepackage{hyperref} 
\usepackage{textcomp}
\usepackage{multirow}
\usepackage{nth}
\usepackage{indentfirst}
\usepackage{siunitx}
\usepackage{changepage}
\usepackage{bm}
\usepackage{setspace}
\usepackage{natbib}
\usepackage{multirow}
\usepackage{multicol}
\usepackage{colortbl}
\usepackage{booktabs}
\usepackage{amsmath}
\usepackage{xcolor}
\usepackage{amsthm}
\theoremstyle{remark}

\usepackage{graphicx}
\usepackage{subcaption}
\usepackage{wrapfig}
\usepackage{enumitem}

\usepackage{minted}

\usepackage{microtype}
\usepackage[normalem]{ulem}
\usepackage{xspace}
\usepackage{tikz}
\usetikzlibrary{pgfplots.groupplots}

\usepackage{pifont}
\usepackage{booktabs}
\usepackage{pgfplots}
\usepackage{pgfplotstable}
\usepackage{amsmath,amsfonts,amsthm,enumitem}
\usepackage{algpseudocode}
\usepackage[linesnumbered,ruled]{algorithm2e}
\usepackage{graphicx}
\usepackage{url}

\usepackage{textcomp}
\usepackage{xcolor}
\usepackage[labelfont=bf]{caption}
\usepackage{subcaption}
\usepackage[T1]{fontenc}
\usepackage[utf8]{inputenc}
\usepackage[most]{tcolorbox}

\DeclareRobustCommand{\github}{\raisebox{-1.5pt}{\includegraphics[height=1.05em]{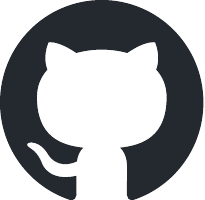}}\xspace}
\DeclareRobustCommand{\webpage}{\raisebox{-1.5pt}
{\includegraphics[height=1.05em]{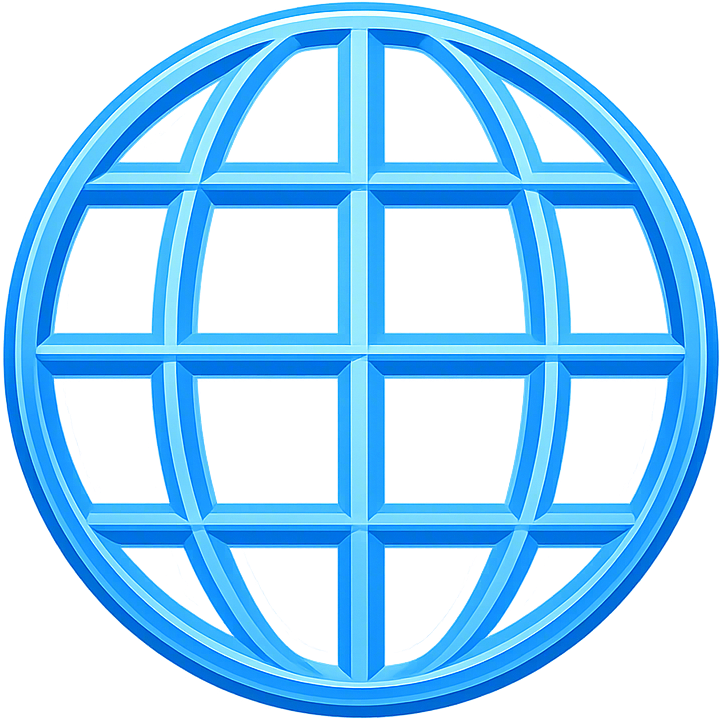}}\xspace}

\usepackage[linesnumbered,ruled]{algorithm2e}
\usepackage{colortbl}
\usepackage{tcolorbox}
\usepackage{listings}
\usepackage{makecell}

\usetikzlibrary{patterns,arrows}
\usepackage{flushend}
\usepackage{balance}
\usepackage{multicol}
\usepackage{multirow}

\usetikzlibrary{trees}

\usepackage[]{mdframed}

\usepackage{pgfplots}
\usetikzlibrary{patterns, pgfplots.groupplots}

\usepackage{xspace}
\usepackage{tkz-kiviat} 
\usetikzlibrary{arrows}

\usepackage{adjustbox}

\definecolor{dkgreen}{rgb}{0,0.6,0}
\definecolor{gray}{rgb}{0.5,0.5,0.5}
\definecolor{mauve}{rgb}{0.58,0,0.82}
\definecolor{orangep}{rgb}{0.71, 0.43, 0.89}
\definecolor{orp}{rgb}{1, 0.7, 0.278}

\definecolor{darkBlue}{rgb}{0.000000,0.000000,0.545098}
\definecolor{darkGreen}{rgb}{0.000000,0.392157,0.000000}
\definecolor{DarkGray}{gray}{0.4}
\definecolor{javared}{rgb}{0.6,0,0} 
\definecolor{javagreen}{rgb}{0.25,0.5,0.35} 
\definecolor{javapurple}{rgb}{0.5,0,0.35} 
\definecolor{javadocblue}{rgb}{0.25,0.35,0.75} 
\definecolor{lightgray}{gray}{0.95}
\definecolor{shadecolor}{RGB}{150,150,150}
\definecolor{blueA}{RGB}{204,229,255}
\definecolor{redA}{RGB}{112,0, 0}
\definecolor{passcolor}{RGB}{144,238,144}
\definecolor{structcolor}{RGB}{255,215,0}
\definecolor{funccolor}{RGB}{255,165,100}
\definecolor{bothcolor}{RGB}{220,80,80}
\definecolor{buildcolor}{RGB}{180,180,180}
\lstdefinestyle{MyCSmallStyle} {
  language=C++,
  frame=none,
  xleftmargin=15pt,
  stepnumber=1, 
  numbers=left, 
  numbersep=5pt,
  numberstyle=\tiny\color[black]{0.177}, 
  belowcaptionskip=\bigskipamount,
  captionpos=b, 
  escapeinside={*'}{'*},
  tabsize=5,
  emphstyle={\bf},
  escapechar=!,
  basicstyle=\scriptsize\ttfamily,
  keywordstyle=\color{javapurple}\bfseries,
  stringstyle=\color{javared},
  commentstyle=\color{javagreen},
  morecomment=[s][\color{javadocblue}]{/**}{*/},
  showspaces=false,
  columns=flexible,
  showstringspaces=false,
  morecomment=[l]{//},
  tabsize=2,
  breaklines=true,
  moredelim=[is][\underbar]{^}{^}
}

\lstdefinestyle{MyJavaSmallStyle} {
  language=Java,
  frame=none,
  xleftmargin=15pt,
  stepnumber=1, 
  numbers=left, 
  numbersep=5pt,
  numberstyle=\color{DarkGray}, 
  belowcaptionskip=\bigskipamount,
  captionpos=b, 
  escapeinside={*'}{'*},
  tabsize=5,
  emphstyle={\bf},
  basicstyle=\scriptsize\ttfamily,
  keywordstyle=\color{javapurple}\bfseries,
  stringstyle=\color{javared},
  commentstyle=\color{javagreen},
  morecomment=[s][\color{javadocblue}]{/**}{*/},
  showspaces=false,
  columns=flexible,
  showstringspaces=false,
  morecomment=[l]{//},
  tabsize=2,
  breaklines=true,
  moredelim=[is][\underbar]{^}{^}
}

\lstdefinelanguage{Scala}{
  keywords={typeof, new, true, false, catch,def,val, function, return, null, catch, switch, var, if, in, while, do, else, case, break, assert, static, void ,declare, const, for, define,fun, ite,class, not, check,sat,String, Int, ArrayList},
  keywordstyle=\color{blue}\bfseries,
  ndkeywords={ export,extends, boolean, throw, implements, import, this, abstract,reduceByKey, reduce, filter, map, reduceByKey, join, Join1, public },
  ndkeywordstyle=\color{mauve}\bfseries,
  otherkeywords={+, =>,<=, ==, >,< , ||},
  identifierstyle=\color{black},
  sensitive=false,
  comment=[l]{//},
  morecomment=[s]{/*}{*/},
  commentstyle=\color{purple}\ttfamily,
  stringstyle=\color{red}\ttfamily,
  morestring=[b]',
  morestring=[b]"
}

\newcommand\todo[1]{\textcolor{red}{TODO: #1}}

\newcommand{\MyPara}[1]{\vspace{.3em}\noindent\textbf{\textit{#1}}~~}

\newcommand{\codefontsmall}[1]{{\small\texttt{#1}}}
\newcommand{\eg}{\emph{e.g.,}\xspace}

\newcommand{\nitr}{{\small{\textsc{NITR}}}\xspace}

\usepackage[most]{tcolorbox}
\definecolor{darkmagenta}{rgb}{0.56, 0.0, 1.0}  

\hypersetup{colorlinks,linkcolor={blue},citecolor={darkmagenta},urlcolor={blue}}

\newtcolorbox{codeblock}{
  colback=gray!10,   
  colframe=gray!50,  
  boxrule=0.5mm,     
  arc=2mm,           
  left=5pt,          
  top=5pt,          
  bottom=5pt,        
}

\newtcbtheorem{proposition}{Proposition}{
  fonttitle=\bfseries,
  fontupper=\normalfont,
  boxrule=0.5pt,
  left=3mm,      
  right=4mm,
  top=2mm,
  bottom=2mm,
  clip upper=false,  
  breakable=false,
  before upper=\setlength{\parindent}{1em}
}{prop}

\newtcbtheorem{definition}{Definition}{
  enhanced,
  colback=cyan!5!white,           
  colframe=cyan!45!blue!60,       
  fonttitle=\bfseries,            
  fontupper=\normalfont,          
  arc=4mm,                        
  boxrule=1pt,
  drop shadow southwest,
  attach boxed title to top left={xshift=6mm, yshift=-2mm},
  boxed title style={
    colback=cyan,
    size=small,
    arc=2mm,                      
    boxrule=0.8pt,
    left=2mm, right=2mm
  },
  before skip=10pt,
  after skip=10pt,
  left=5mm,
  right=5mm,
  top=3mm,
  bottom=3mm
}{cor} 

\renewcommand{\ghlink}{https://github.com/alibaba/ROLL}

\newcommand*\justify{%
  \fontdimen2\font=0.4em
  \fontdimen3\font=0.2em
  \fontdimen4\font=0.1em
  \fontdimen7\font=0.1em
  \hyphenchar\font=`\-
}

\renewcommand{\texttt}[1]{%
  \begingroup
  \ttfamily
  \begingroup\lccode`~=`/\lowercase{\endgroup\def~}{/\discretionary{}{}{}}%
  \begingroup\lccode`~=`[\lowercase{\endgroup\def~}{[\discretionary{}{}{}}%
  \begingroup\lccode`~=`.\lowercase{\endgroup\def~}{.\discretionary{}{}{}}%
  \catcode`/=\active\catcode`[=\active\catcode`.=\active
  \justify\scantokens{#1\noexpand}%
  \endgroup
}

\definecolor{blueviolet}{RGB}{138,43,226}

\newtcolorbox{abstractbox}{
    colback=blue!5!white,     
    frame empty,              
    boxrule=1pt,              
    arc=4mm,                  
    left=8pt,                 
    right=8pt,                
    top=8pt,                  
    bottom=8pt,                
    opacityback=0.9
}

\title{Needle in the Repo: A Benchmark for Maintainability in AI-Generated Repository Edits}

\author{
{\bf 
\mbox{Haichao Zhu$^{2,\dagger}$, Qian Zhang$^{1,\dagger}$, 
Jiyuan Wang$^{3}$, 
Zhaorui Yang$^{1}$, 
Yuxin Qiu$^{1}$} \\
\mbox{
$^1$UC Riverside \quad
$^2$ Independent Researcher \quad
$^3$Tulane University \quad
}
}
}

\definecolor{cyan}{cmyk}{.3,0,0,0}

\begin{document}

\maketitle
\renewcommand{\thefootnote}{}
\footnotetext{$^\dagger$Equal contribution.}
\footnotetext{$^\ast$Corresponding authors: \url{lszhuhaichao@gmail.com}; \url{qzhang@cs.ucr.edu}.}
\renewcommand{\thefootnote}{\arabic{footnote}}

\vspace{-4mm}
\begin{abstractbox}
\begin{center}
\vspace{-1mm}
\textbf{\Large Abstract}
\end{center}

AI coding agents can now complete complex programming tasks, but existing evaluations largely emphasize behavioral correctness and often overlook maintainability risks such as weak modularity or testability. Automation without accountability shifts invisible cost downstream because hidden maintainability failures incurred today are deferred to the developers who must extend, debug, and sustain these systems tomorrow, raising significant technical and organizational concerns for software engineering. We present {Needle in the Repo} (\textsc{NITR}), a diagnostic, probe-and-oracle framework for evaluating whether behaviorally correct repository edits preserve maintainable structure. The key idea is to distill recurring software engineering wisdom into \emph{controlled probes} embedded in small, realistic multi-file codebases, with each probe designed so that success depends primarily on one targeted maintainability dimension. 
Each probe is paired with a hidden evaluation harness that combines (i) functional tests for required behavior, and (ii) structural oracles that encode the targeted maintainability constraint and return interpretable diagnoses.

Using \textsc{NITR}, we extensively evaluate 23 coding configurations across GPT, Claude, Gemini, and Qwen families in both direct-inference and agent-based settings.
We find that current AI coding systems remain far from robust. On average, all AI coding configurations solve only 36.2\% of cases, the best reaches 57.1\%, and performance drops from 53.5\% on micro cases to 20.6\% on multi-step cases.
The hardest pressures are architectural rather than local edits, especially dependency control (4.3\%) and responsibility decomposition (15.2\%).
Moreover, 64/483 outcomes (13.3\%) pass all functional tests yet fail the structural oracle.
Under our harness, agent-mode configurations improve average performance from 28.2\% to 45.0\%, but does not eliminate the architectural failures. These results show that progress in code generation is not yet progress in maintainable code evolution, and that \textsc{NITR} exposes a critical failure surface largely missed by conventional evaluation.

\begin{center}
\vspace{-1mm}
  \github\ \href{https://github.com/ucr-riple/NITR}{Repo}\hspace{1.5em}
  \webpage\ \href{https://www.cs.ucr.edu/~qzhang/nitr.html}{WebPage}\hspace{1.5em}
\end{center}

\end{abstractbox}

\begin{figure*}[h]
\centering
\includegraphics[width=1\textwidth]{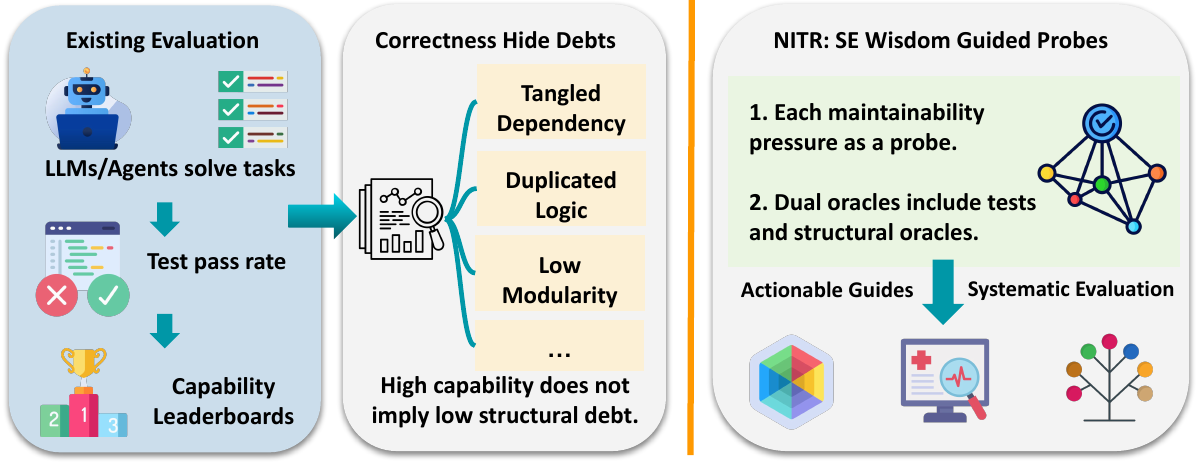}
\caption{Existing evaluations rank models by behavioral success, but leaderboard performance says little about maintainability risk. \textsc{NITR} uses curated probes and structural oracles to expose failure modes beyond test passing.
}
\label{fig:teaser}
\end{figure*}

\section{Introduction}\label{sec:intro}

Artificial intelligence (AI) is rapidly reshaping software development, demonstrating strong capabilities in code understanding, generation, execution, and iterative debugging~\cite{jimenez2024swebench,yang2024sweagent,wang2025openhands}. We use the term \emph{coding agents} to refer to large language model (LLM)-based systems equipped with scaffolding such as repository navigation~\cite{yang2024sweagent,ouyang2025repograph}, code editing~\cite{yang2024sweagent,wang2025openhands}, program execution~\cite{wang2024codeact}, and feedback-driven repair~\cite{shinn2023reflexion,bouzenia2025repairagent}. These agents now operate directly inside real development environments, making their evaluation increasingly important for understanding what kinds of software engineering (SE) work they can actually support.

\vspace{.3em}
\noindent{\bf The Gap: Completion is Not Construction.}
Recent benchmarks have made substantial progress by moving beyond single-file completion toward repository-level tasks~\cite{liu2023repobench,jimenez2024swebench,li2024evocodebench}. For example, SWE-bench and its variants~\cite{jimenez2024swebench,yang2025swebench} evaluate LLMs on issue resolution across 12 GitHub repositories.
However, these evaluations still define success primarily in terms of \emph{behavioral correctness}, operationalized as pass/fail performance on predefined test cases. This leaves a fundamental gap between what current benchmarks measure and what SE practice actually requires, illustrated in Figure~\ref{fig:teaser}.

First, test pass/fail is a narrow view of software quality. A patch may satisfy the target tests while still duplicating logic, bypassing an existing abstraction, or spreading a small feature across unrelated modules, thereby making the repository harder to extend and maintain over time. Recent studies of non-functional quality in LLM-generated code likewise find that behavioral success does not guarantee maintainability, reliability, or broader quality attributes~\cite{sun2025qualityassurance,molison2025maintainability}. This concern also appears in practitioner discussion. For example, one recent Reddit comment~\cite{reddit-post} observed that “The real failure mode I keep seeing is maintenance.”

Second, test-centric evaluations provide weak diagnostic insight. They indicate whether agents completed a task, but not which maintainability capabilities they handled well or failed to preserve. Previous work has shown that correctness-based benchmarks often miss specific failure modes and overlook code-quality concerns that matter in practice~\cite{hu2024fauneval,zheng2024beyondcorrectness,chen2026sweci}. As a result, they rank models by outcome, but reveal little about where current LLMs still fall short.

\vspace{.3em}
\noindent{\bf This Work.}
This work addresses a core limitation of current code-agent evaluation: passing repository tasks does not tell us whether the resulting code remains maintainable, or which maintainability capabilities current LLMs fail to preserve. We present {\sc Needle in the Repo} (\textsc{NITR}), a probe-and-oracle framework for evaluating maintainability preservation in AI-generated repository edits. Throughout this paper, we use \emph{maintainability failure} to refer to code that is behaviorally correct for the current task but makes future evolution harder to extend, test, or reason about safely.

Our central insight is that software engineering wisdom can be distilled into small, repository-grounded probes with explicit design boundaries and executable structural checks. \textsc{NITR} operationalizes 9 maintainability dimensions through targeted C++ repository probes: Change Locality, Reuse and Repo Awareness, Responsibility Decomposition, Extension Structure, Interface and Substitutability Discipline, Dependency Control, Testability and Determinism, State Ownership and Lifecycle, and Side-Effect Isolation.

{\sc NITR} comprises 21 curated C++ repository probes spanning nine maintainability dimensions. Each case encodes one primary repository-evolution pressure through a natural multi-file change request and is paired with hidden functional tests and structural probes, enabling diagnosis beyond test passing.




\vspace{.3em}
\noindent{\bf Key Findings.}
Using \textsc{NITR}, we evaluate 23 coding configurations spanning GPT, Claude, Gemini, and Qwen families in both direct-inference and agent-mode settings.

We present, to our knowledge, the first probe-and-oracle study focused on maintainability preservation in AI-generated repository edits and report three findings.
First, current coding systems remain far from robust at maintainability-preserving repository evolution: the average configuration solves only 36.2\% of cases, the best reaches 57.1\%, and five probes are unsolved by all configurations.
Second, current systems struggle most not with isolated edits, but with changes that must preserve deeper repository structure, including dependency control (4.3\% pass rate), responsibility decomposition (15.2\%), and abstraction-respecting extension (26.1\%).
Third, test passing is an unreliable proxy for maintainability: 64/483 outcomes (13.3\%) are behaviorally correct yet structurally wrong, and although agent-mode configurations under our harness raises average performance from 28.2\% to 45.0\%, it does not eliminate these core failures.
Together, these results show that \textsc{NITR} reveals hidden structural debt that conventional evaluations largely miss. 

This work makes three contributions:
\begin{itemize}[leftmargin=1.2em]
  \item \textsc{NITR} turns maintainability-preserving repository edit evaluation into a controlled object of study by pairing curated repository probes with hidden structural oracles.
  \item We provide the first probe-and-oracle empirical study of maintainability preservation in AI-generated repository edits, showing that test-passing often misses structural failure and that the hardest pressures are architectural rather than local.
  \item We release \textsc{NITR} as an open-source suite to support future research on maintainability-aware coding agents and repository-level software engineering evaluation.
\end{itemize}

NITR is not intended as a universal taxonomy or a naturalistic sample of all software tasks; it is a diagnostic suite designed to make specific maintainability pressures observable, executable, and comparable. In other words, it does not attempt to cover all maintainability phenomena, but to make a set of recurring, practically important repository-evolution pressures precise enough to evaluate, compare, and analyze.



\section{Motivating Example}\label{sec:motivation}

Coding agents are often evaluated as if passing tests were sufficient evidence of code quality. The motivating example in case~001 of \textsc{NITR} shows why this assumption is too weak. In this task, a small utility function \codefontsmall{add} is widely used across a codebase. The agent is asked, over three incremental steps, to extend it to support additional numeric types: first \codefontsmall{int}, then \codefontsmall{float} and \codefontsmall{double}, and finally \codefontsmall{long long}. Throughout all steps, the call sites in \codefontsmall{app/main.cc} remain fixed and must not be modified.

Listing~\ref{lst:motivating} shows two representative implementations submitted by agents.
The \emph{overloaded} implementation (lines 2--8) is the approach taken by 17 of 23
evaluated configurations.
It introduces a separate, copy-pasted definition for every numeric type.
Functionally, both implementations are equivalent: all unit tests pass.
The tests check correct addition output for each required type; they do not
inspect the structure of the implementation.

The structural probe in \textsc{NITR}, however, evaluates a different criterion.
It checks that the implementation provides a \emph{centralized reusable core}
(detected via generic-function patterns such as a template) and that no more than
one explicit per-type overload definition is present.
The overloaded implementation (lines 2--8) violates both conditions and is
therefore marked as failing, even though every functional test passes.

\setminted{
  fontsize=\footnotesize,
  linenos,
  numbersep=6pt,
  xleftmargin=1.5em,
  frame=lines,
  framesep=2mm,
}

\begin{listing}[tbp]
\caption{Two implementations of a multi-type \texttt{add} function. Both pass all unit
tests, but the overloaded version (top) fails the structural probe because it
introduces a new definition for each type added, amplifying change across the codebase.
The template version (bottom) passes both unit tests and the structural probe. }
\label{lst:motivating}
\begin{minted}[fontsize=\footnotesize, linenos]{cpp}
// Non-maintainable: submitted by 17/23 agents. Each new type
// requires duplicating a definition (change amplification).
namespace solid::case001 {
  int       add(int a,       int b)       { return a + b; }
  float     add(float a,     float b)     { return a + b; }
  double    add(double a,    double b)    { return a + b; }
  long long add(long long a, long long b) { return a + b; }
}

// Maintainable: a single generic definition handles all types.
// Adding support for a new type requires zero code change.
namespace solid::case001 {
  template <typename T>
  T add(T a, T b) { return a + b; }
}
\end{minted}
\end{listing}

This failure has direct maintenance consequences. If the codebase later needs to support
\codefontsmall{unsigned int} or \codefontsmall{float128}, the overloaded design requires adding another specialized definition and modifying the library interface. In a multi-step tasks, the problem becomes worse. Once an agent has introduced the overload pattern, later steps may continue extending that pattern rather than repairing the abstraction, compounding the original design mistake.

This example illustrates the exact failure mode that \textsc{NITR} is designed to expose: code can be behaviorally correct yet structurally poor to evolve. 


\section{Maintainability Probe Design}\label{sec:approach}

\begin{figure}[t]
  \centering
  \includegraphics[width=0.85\columnwidth]{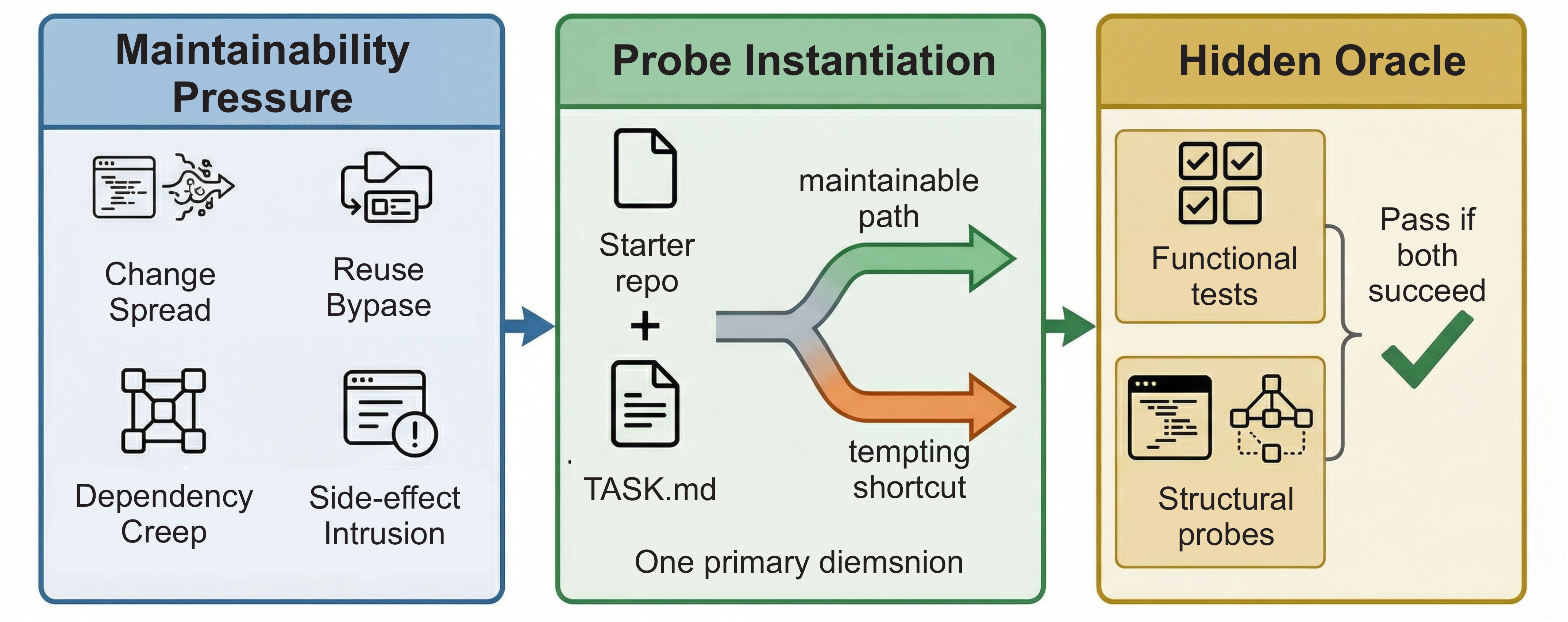}
  \caption{
  From maintainability pressure to diagnostic probe.
  \textsc{NITR} summarizes SE practices into 9 recurring repository-evolution pressures, instantiates each target pressure as a compact probe with starter code and an agent-facing task, and pairs the probe with functional and structural oracles.
  }
  \label{fig:probe-pipeline}
\end{figure}

We construct \textsc{NITR} in three parts as shown in Figure~\ref{fig:probe-pipeline}.
First, we define a maintainability design space grounded in recurring software-evolution pressures (Section~\ref{sec:approach-1}).
Second, we instantiate each target pressure as a compact repository probe with starter code and an agent-facing task (Section~\ref{sec:approach-2}).
Third, we package each probe with hidden functional tests and structural oracles that distinguish maintainable solutions from tempting shortcut solutions (Section~\ref{sec:approach-3}).

\textsc{NITR} is organized as a collection of expert-curated maintainability probes rather than generic coding tasks mined from GitHub that may already be familiar to LLMs.
Each probe is a small repository-level scenario designed to expose one primary maintainability pressure under realistic software evolution.
The evaluation target is whether AI's chosen implementation strategy preserves maintainable structure. The resulting suite contains 21 C++ probes spanning 9 software engineering dimensions, including 10 micro probes and 11 multi-step probes detailed below.

\subsection{From Principles to Maintainability Dimensions}\label{sec:approach-1}

Our starting point is that software engineering wisdom already provides a practical basis for judging whether code will remain easy to extend, modify, test, and integrate over time. Accordingly, \textsc{NITR} begins from the design principles and engineering practices that developers use to assess code quality in real repositories.

To define the construction design space of \textsc{NITR}, we reviewed established software engineering principles, including SOLID~\cite{martin2003agile}, together with maintainability practices such as code reuse~\cite{haefliger2008code,krueger1992software} and testability~\cite{freedman1991testability,parker1982design}.

From this author-driven consolidation, we derived the nine dimensions by consolidating recurrent repository-evolution pressures rather than by starting from a fixed taxonomy alone. Concretely, we first enumerated common structural failure pressures that arise during repository change, such as change amplification, duplicate-path introduction, responsibility leakage, dependency creep, test-seam erosion, lifecycle scattering, and side-effect intrusion. We then grouped these pressures into broader, reusable construction axes, using classical software-design principles such as SOLID as an organizing lens where appropriate. The resulting nine dimensions are therefore best understood as probe-construction axes for maintainability-relevant repository evolution, not as an exhaustive ontology of maintainability.

The nine dimensions include:
(D1) \emph{Change Locality}, whether a requested change remains localized rather than forcing scattered edits;
(D2) \emph{Reuse and Repo Awareness}, whether the solution reuses and adapts existing repository logic instead of re-implementing it in parallel;
(D3) \emph{Responsibility Decomposition}, whether distinct concerns remain separated rather than collapsing into one component; 
(D4) \emph{Extension Structure}, whether new behavior is added through clean extension rather than ad hoc special cases;
(D5) \emph{Interface and Substitutability Discipline}, whether interfaces remain narrow, coherent, and safe for interchangeable implementations;
(D6) \emph{Dependency Control}, whether modules avoid unnecessary coupling and depend only on what they need;
(D7) \emph{Testability and Determinism}, whether the design preserves clean test seams and avoids hidden nondeterminism; 
(D8) \emph{State Ownership and Lifecycle}, whether mutable state has clear ownership and lifetime boundaries and
(D9) \emph{Side-Effect Isolation}, whether effects such as I/O, logging, and global mutation are kept separate from core logic. Table 1 reports coverage at the dimension level, while the design matrix records the concrete failure pressures each dimension was designed to capture.


Each case declares exactly one \texttt{primary\_dimension}, defined as the main maintainability pressure the case is designed to isolate. Optional \texttt{secondary\_dimensions} record supporting pressures when needed. Assigning one primary dimension keeps each probe interpretable and supports balanced coverage across the design space, while still allowing secondary pressures to appear as context.

\subsection{Instantiating Maintainability Probes}\label{sec:approach-2}

\begin{wrapfigure}{r}{0.56\textwidth}
  \centering
  \includegraphics[width=0.56\columnwidth]{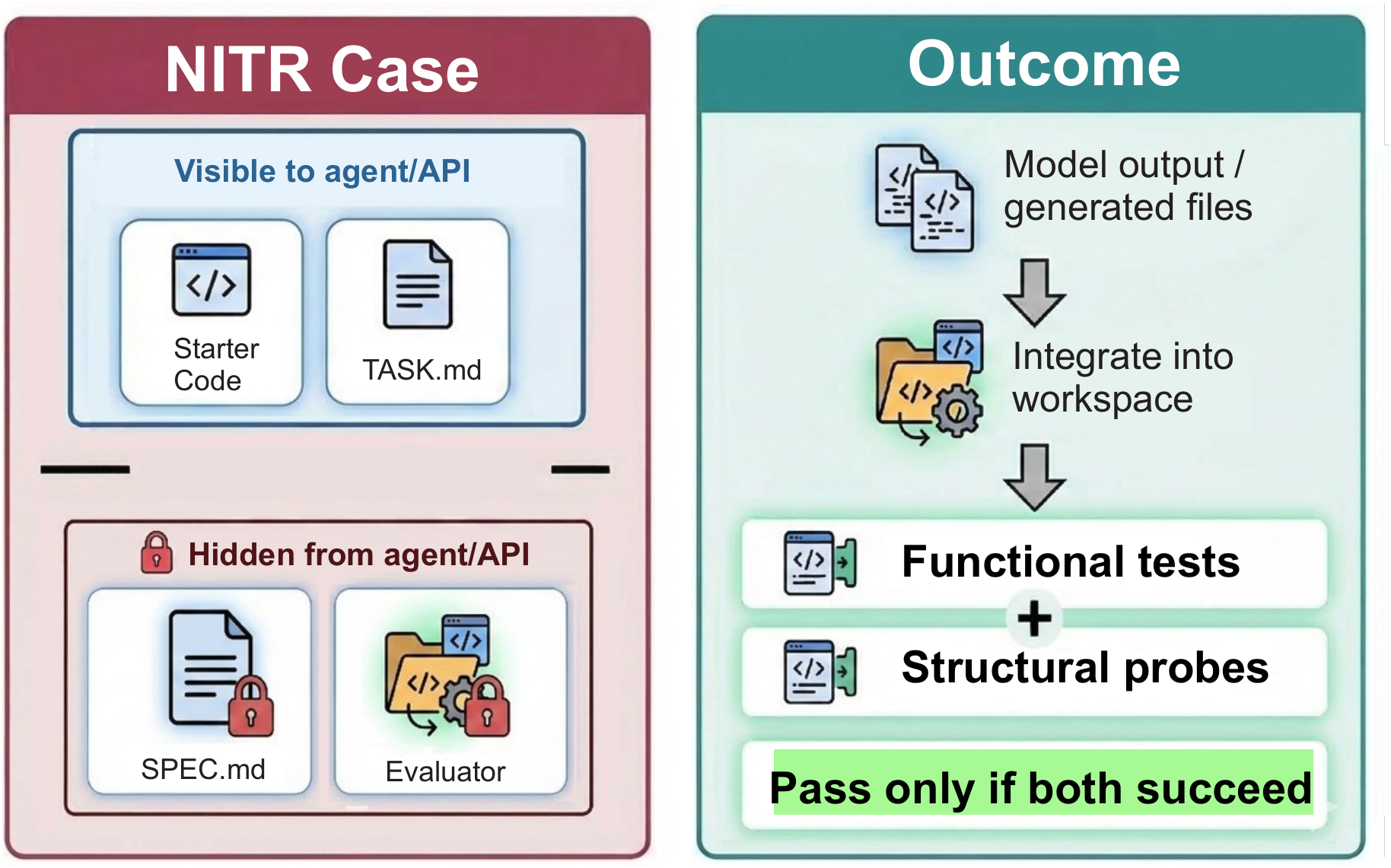}
  \caption{
Anatomy of a maintainability probe.
Each probe contains starter code, an agent-visible task (\texttt{TASK.md}), an author-facing specification (\texttt{SPEC.md}), and a hidden evaluator.
During evaluation, the model sees only the starter code and \texttt{TASK.md}; the probe passes only if the generated edits satisfy both the functional tests and the structural probes.
  }
  \label{fig:case-anatomy}
\end{wrapfigure}

Given a target dimension, we instantiate it as a small, self-contained C++ repository probe. A probe is a jointly designed artifact consisting of a repository scenario, starter code, an agent-visible task, and hidden evaluation logic, as illustrated in Figure~\ref{fig:case-anatomy}. Each probe is constructed so that the target maintainability pressure arises naturally, a shortcut solution remains plausible, and the difference between the two can be diagnosed automatically. Please note that this curation is deliberate rather than incidental. In mined tasks, the intended design boundary is often implicit, underspecified, or entangled with unrelated repository history, making maintainability judgment difficult to operationalize. NITR instead uses compact, authored repository scenarios where the maintainable path and the tempting shortcut are both behaviorally plausible, but structurally distinguishable. This enables interpretable diagnosis of edit strategy rather than only outcome success.

\vspace{.3em}
\noindent{\bf Step 1: Repository Scenario Selection.}
We first select a compact engineering change that naturally exposes the target maintainability pressure. The scenario should be realistic and narrow in scope, while still admitting two plausible implementation paths: one that preserves repository structure and one that satisfies the required behavior through a shortcut.

For example, the curated Case~021 targets D2: \emph{Reuse and Repo Awareness}. The requested change is to add support for an inline filter form such as \texttt{status=open,priority=3}. The repository already supports filtering through an existing structured pipeline, so the core question is whether the new inline entry point reuses that pipeline or introduces a second parsing-and-validation path. Listing~\ref{lst:case021-oracle} shows the associated structural oracle.

\vspace{.3em}
\noindent{\bf Step 2: Starter-Code Shaping.}
We then shape the starter repository so that the intended maintainable path already exists in the codebase. In Case~021, the starter code already contains a repository-native filter representation, an existing parse/validate pipeline, and a designated validation module. This creates two clear solution paths. A maintainable solution parses the inline string only far enough to construct the existing internal representation and then delegates to the current pipeline. A shortcut solution instead introduces inline-only logic for numeric parsing, field interpretation, or error classification. These shortcut patterns are encoded directly in the hidden oracle checks in Listing~\ref{lst:case021-oracle}, including checks for \texttt{std::stoi}, \texttt{std::isdigit}, duplicated field literals, and duplicated error literals.


\begin{figure}[t]
  \centering
  \includegraphics[width=0.6\columnwidth]{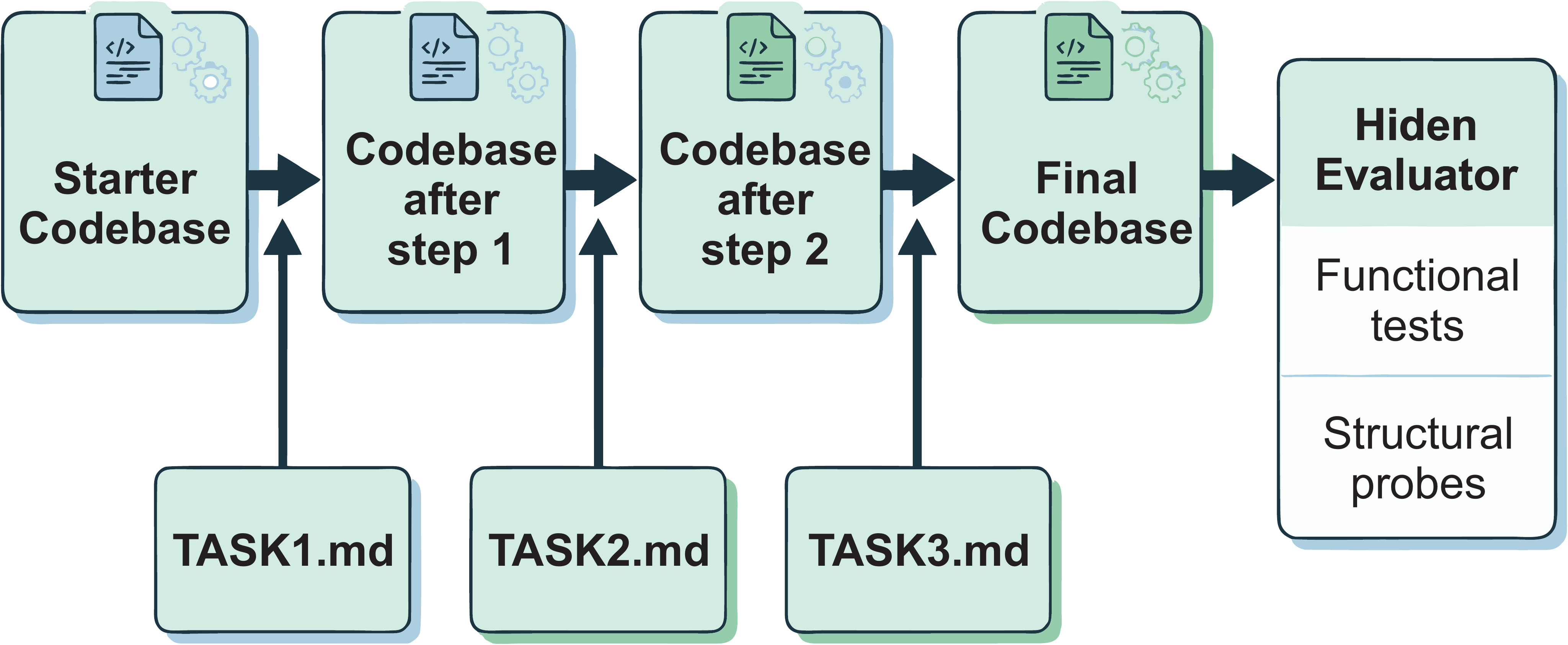}
  \caption{
Multi-step probe execution.
Each step is applied to the codebase produced by the previous step, so early design choices persist across later changes.
The final codebase is then evaluated with tests and structural oracles.
  }
  \label{fig:multi-step}
\end{figure}

\vspace{.3em}
\noindent{\bf Step 3: Agent-Facing Task Design.}
For each probe, we write a concise \texttt{TASK.md} as a natural engineering request. It specifies the required behavior, but does not reveal the maintainability dimension, the intended design move, or the hidden oracle logic.

In Case~021, the visible task asks the agent to add support for the inline filter form while preserving existing behavior. It does not say ``reuse the validation module,'' does not mention D2, and does not warn against introducing a helper such as \texttt{ParseInlineFilter} with its own integer parsing and field checks. 


\vspace{.3em}
\noindent{\bf Step 4: Micro versus Multi-Step Probes.}
We use a micro probe when the targeted maintainability pressure can be exposed by a single localized change, and a multi-step probe when it only becomes visible under continued evolution. In a multi-step probe, the model receives a sequence of task files, each applied to the code produced at the previous step, as illustrated in Figure~\ref{fig:multi-step}. This makes early design decisions persistent across later changes.

Case~021 is a micro probe because the D2 failure is visible after one change. Once the agent adds the inline entry point, the solution either routes it through the existing pipeline or introduces inline-only parsing and validation logic. The hidden oracle can detect that shortcut immediately through patterns such as \texttt{std::stoi}, \texttt{std::isdigit}, duplicated field handling, or duplicated error classification, so no further evolution is needed to expose the difference. By contrast, dimensions such as \emph{Dependency Control} or \emph{Responsibility Decomposition} often require multi-step probes, because a design that looks acceptable at step 1 may only fail later when added features force one component to absorb logging, provider-specific logic, or extra dependencies.

\begin{listing}[t]
\caption{Excerpt of the hidden structural oracle for Case~021. The individual string-level checks shown here are not used as standalone maintainability judgments; they function only as case-specific signals within a jointly designed repository context.}
\label{lst:case021-oracle}
\begin{minted}[fontsize=\footnotesize, linenos]{python}
# Check 1: The inline entrypoint should not perform its own
# numeric parsing. In the intended design, integer validation 
# stays in the existing filter-validation path rather than being  
# reimplemented inline.
has_stoi = "std::stoi" in content
has_isdigit = "std::isdigit" in content

if has_stoi or has_isdigit:
    fail("Suspicious duplicate numeric validation")

# Check 2: Repeating field and error literals outside the
# designated validation files often signals a shadow parser with
# its own lookup tables and error mapping logic.
has_inline_parser = "ParseInlineFilter" in content
field_hits = sum(literal in content for literal in FIELD_LITERALS)
error_hits = sum(literal in content for literal in ERROR_LITERALS)

if has_inline_parser and field_hits > 0 and error_hits > 1:
    fail("Suspicious duplicated error classification")
\end{minted}
\end{listing}

\subsection{Packaging Dual Oracles}\label{sec:approach-3}

Each probe is paired with two hidden evaluators: \emph{functional tests}, which check the requested behavior, and \emph{structural oracles}, which check whether the implementation preserves the intended maintainability constraint. A probe passes only if both succeed.

The structural oracle is probe-specific, but its design follows recurring patterns aligned with the nine dimensions. For \emph{Change Locality} (D1), it checks whether a small requirement change remains localized. For \emph{Reuse and Repo Awareness} (D2), it checks reuse of existing repository affordances and rejects duplicate implementation. For \emph{Responsibility Decomposition} (D3), \emph{Extension Structure} (D4), and \emph{Dependency Control} (D6), it checks whether new behavior stays within the intended module boundaries rather than expanding central dispatch logic or introducing unnecessary coupling. For \emph{Interface and Substitutability Discipline} (D5), it checks that interfaces remain narrow and that callers do not branch on concrete subtype identity. For \emph{Testability and Determinism} (D7), it flags hidden wall-clock or randomness usage and verifies that the intended deterministic seam is preserved. For \emph{State Ownership and Lifecycle} (D8), it checks that lifecycle-sensitive state transitions remain within the designated ownership boundary. For \emph{Side-Effect Isolation} (D9), it ensures that logging, tracing, or explanation paths do not intrude into core decision logic. These checks are lightweight and targeted rather than full semantic program analyses, but they are sufficient to distinguish the intended maintainable solution family from common shortcut patterns.

Case 021 illustrates the oracle pattern. Functional tests alone cannot distinguish reuse of the existing filter pipeline from a shadow parser, so the structural oracle checks for duplicated parsing and validation outside the designated repository path. Listing~\ref{lst:case021-oracle} shows an excerpt: it flags uses of \texttt{std::stoi} or \texttt{std::isdigit}, rejects inline-only integer handling when \texttt{FilterValueKind::kInteger} appears together with \texttt{ParseInlineFilter}, and detects repeated field or error logic outside the existing validation flow. In this way, the oracle distinguishes reuse of repository affordances from parallel reimplementation.

\vspace{.3em}
\noindent{\bf Oracle validation.} During case construction, each oracle was validated against intended maintainable solutions, representative shortcut solutions, and near-miss variants. We iterated until the oracle stably separated the intended solution family from targeted shortcuts under small implementation variation; representative intended, shortcut, and near-miss variants are included in the artifact.


\begin{wraptable}{r}{0.6\textwidth}
\centering
\caption{The nine maintainability dimensions used as probe-construction axes in \textsc{NITR}. Each case is assigned exactly one primary dimension. The last column reports how many cases in that dimension involve multiple steps.
}
\label{tab:dimensions}
\scalebox{0.9}{
\begin{tabular}{llcc}
\toprule
\textbf{Dim.} & \textbf{Name} & \textbf{\#Cases} & \textbf{\#Multi-step} \\
\midrule
D1 & Change Locality & 3 & 1 \\
D2 & Reuse \& Repo Awareness & 3 & 2 \\
D3 & Responsibility Decomposition & 2 & 1 \\
D4 & Extension Structure & 2 & 1 \\
D5 & Interface \& Substitutability & 2 & 1 \\
D6 & Dependency Control & 2 & 1 \\
D7 & Testability \& Determinism & 3 & 2 \\
D8 & State Ownership \& Lifecycle & 2 & 1 \\
D9 & Side-Effect Isolation & 2 & 1 \\
\midrule
   & \textbf{Total} & \textbf{21} & \textbf{11} \\
\bottomrule
\end{tabular}
}
\end{wraptable}

\begin{table*}[t]
\centering
\caption{All 21 repositories in \nitr{} with their dimension, granularity, and aggregate
outcomes across 23 evaluated configurations.
\textbf{S} counts are bolded where $\geq 4$, highlighting probes where functionally correct code still fails the maintainability oracle.}
\label{tab:cases}
\resizebox{\textwidth}{!}{%
\setlength{\tabcolsep}{5pt}
\begin{tabular}{l*{21}{c}}
\toprule
 & \rotatebox{90}{\textbf{001}}
 & \rotatebox{90}{\textbf{002}}
 & \rotatebox{90}{\textbf{003}}
 & \rotatebox{90}{\textbf{004}}
 & \rotatebox{90}{\textbf{005}}
 & \rotatebox{90}{\textbf{006}}
 & \rotatebox{90}{\textbf{007}}
 & \rotatebox{90}{\textbf{008}}
 & \rotatebox{90}{\textbf{009}}
 & \rotatebox{90}{\textbf{010}}
 & \rotatebox{90}{\textbf{011}}
 & \rotatebox{90}{\textbf{012}}
 & \rotatebox{90}{\textbf{013}}
 & \rotatebox{90}{\textbf{014}}
 & \rotatebox{90}{\textbf{015}}
 & \rotatebox{90}{\textbf{016}}
 & \rotatebox{90}{\textbf{017}}
 & \rotatebox{90}{\textbf{018}}
 & \rotatebox{90}{\textbf{019}}
 & \rotatebox{90}{\textbf{020}}
 & \rotatebox{90}{\textbf{021}} \\
\midrule
Dim.
 & D1 & D2 & D2 & D3 & D4
 & D5 & D5 & D6 & D7 & D9
 & D1 & D8 & D1 & D4 & D6
 & D7 & D8 & D7 & D9 & D3
 & D2 \\
Type
 & \emph{m} & \emph{m} & \emph{m} & \emph{m} & \emph{m}
 & \emph{$\mu$} & \emph{m} & \emph{$\mu$} & \emph{$\mu$} & \emph{$\mu$}
 & \emph{$\mu$} & \emph{$\mu$} & \emph{$\mu$} & \emph{$\mu$} & \emph{m}
 & \emph{m} & \emph{m} & \emph{m} & \emph{m} & \emph{$\mu$}
 & \emph{$\mu$} \\
Pass
 &  5 &  0 & 22 &  0 &  0
 &  2 & 10 &  2 & 21 & 21
 & 10 & 23 & 18 & 19 &  0
 &  8 &  4 &  2 &  1 &  7
 &  0 \\
\textbf{S}
 & \textbf{13} & 0 & 0 & 0 & 0
 & \textbf{4}  & 0 & \textbf{4} & 0 & 0
 & 0 & 0 & 0 & 0 & 0
 & 0 & 0 & 0 & 0 & \textbf{11}
 & 0 \\
\bottomrule
\end{tabular}%
}
\par\smallskip
\noindent
\vspace{.3em}
\emph{m} = multi-step;\quad $\mu$ = micro;\quad
\textbf{S} = functional tests pass, but the maintainability oracle fails.
\vspace{.3em}
\end{table*}

\subsection{Evaluation Protocol}

For each probe, the harness creates a temporary workspace, applies the model-generated file edits, configures and builds the project with \texttt{CMake}, runs the hidden functional tests, and then executes the structural oracles. A probe is marked \emph{pass} only if both the functional and structural evaluations succeed.

This strict conjunction is intentional. \textsc{NITR} is designed to expose a failure surface that ordinary test-centric evaluation misses: solutions that satisfy the requested behavior while introducing structural shortcuts. Table~\ref{tab:dimensions} summarizes the nine maintainability dimensions covered by \textsc{NITR}, and Table~\ref{tab:cases} provides the full probe inventory, including each case's dimension, granularity, and aggregate outcome counts across the evaluated configurations.

\section{Experimental Setup}\label{sec:setup}

We use {\sc NITR} to study coding agents and answer questions:

\begin{itemize}[leftmargin=1.5em]
\item \textbf{RQ1:} How well do contemporary coding systems perform on the \textsc{NITR} case suite?
\item \textbf{RQ2:} Which maintainability dimensions are most challenging for current coding systems?
\item \textbf{RQ3:} What structural failure patterns do current AI coding tools exhibit, and to what extent are these failures missed by functional tests alone?
\item \textbf{RQ4:} Under our constrained evaluation harness, to what extent do agent-mode configurations outperform direct API-based configurations from the same model family on maintainability preservation?
\end{itemize}

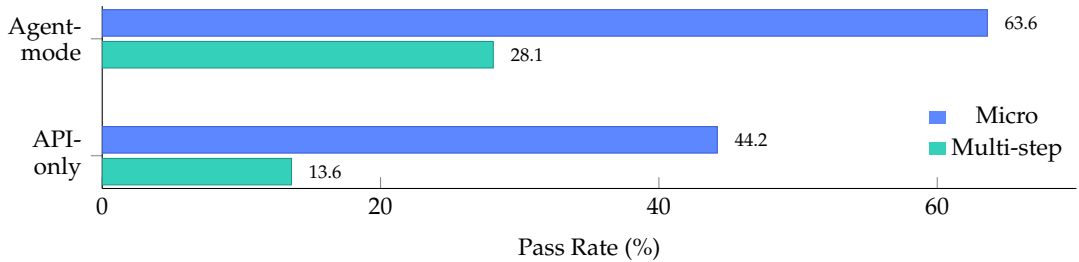
\begin{figure}[t]
\centering

\begin{tikzpicture}
\vspace{.5em}
\begin{axis}[
    xbar,
    width=0.9*\linewidth,
    height=4.0cm,
    xmin=0,
    xmax=70,
    bar width=10pt,
    axis x line*=bottom,
    axis y line*=left,
    symbolic y coords={Agent-mode,API-only},
    ytick=data,
    yticklabels={{API-\\only},{Agent-\\mode}},
    y dir=reverse,
    enlarge y limits=0.28,
    yticklabel style={xshift=0pt},
    xlabel={Pass Rate (\%)},
    xtick={0,20,40,60},
    xticklabels={0,20,40,60},
    legend style={
      at={(1.00,0.10)},
      anchor=south east,
      draw=none,
      fill=none,
      font=\small,
    },
    legend columns=1,
    reverse legend,
    legend image code/.code={
      \draw[#1, draw=none] (0cm,-0.08cm) rectangle (0.22cm,0.08cm);
    },
    tick label style={font=\small, align=right},
    label style={font=\small},
    nodes near coords,
    every node near coord/.append style={font=\scriptsize, black, anchor=west, xshift=3pt},
    point meta=explicit symbolic,
]
\addplot+[fill={rgb,255:red,53; green,208; blue,186}, draw={rgb,255:red,28; green,150; blue,135}] coordinates {
   (13.6,API-only) [13.6]
    (28.1,Agent-mode) [28.1]
};
\addplot+[fill={rgb,255:red,93; green,135; blue,255}, draw={rgb,255:red,60; green,95; blue,210}] coordinates {
(44.2,API-only) [44.2]
    (63.6,Agent-mode) [63.6]
};
\legend{Multi-step,Micro}
\end{axis}
\end{tikzpicture}
\caption{Both API-only and agent-mode systems perform substantially better on micro cases than on multi-step cases, indicating that the main difficulty lies in evolutionary tasks rather than isolated edits.}
\label{fig:granularity_passrate}
\end{figure}

\vspace{.3em}
\noindent{\bf Models and Agents.}
We evaluate 23 configurations from multiple providers: 11 agent-mode and 12 API-only.
Agent-mode systems are accessed through provider-managed CLI surfaces that provide agentic scaffolding, such as multi-turn interaction and managed execution environments.
API-only systems are accessed through direct model endpoints without such scaffolding.
For API-mode evaluation, Qwen, Gemini, and Claude are served through Google Cloud Vertex AI~\cite{vertex}, while OpenAI models are accessed through the official OpenAI API~\cite{openaiapi}.

\vspace{.3em}
\noindent{\bf Submission Format and Execution Setup.}
To standardize outputs, all systems must return edits as a JSON dictionary mapping file names to file contents.
A local Python harness materializes each submission by overwriting the corresponding repository files.
In multi-step cases, agent-mode systems receive tasks sequentially and may observe their own prior edits, whereas API-only systems receive each step independently together with the relevant prior context.
In all settings, models see only the agent-facing \codefontsmall{TASK.md}, without hidden hints or additional maintainability instructions.

\vspace{.3em}
\noindent{\bf Experimental Constraints.}
To prevent evaluator leakage, neither agent-mode nor API-only systems can access the hidden unit tests or Python evaluation scripts.
Agent-mode systems are additionally restricted to read-only repository access: they may inspect files in the target repository, but cannot execute arbitrary commands, use external tools, or access files outside the workspace.
These controls ensure that outputs are derived only from \codefontsmall{TASK.md} and the visible repository contents.

\section{Experimental Results and Findings}\label{sec:results}

\newcommand{\pnone}[1]{\cellcolor{red!18}{\small #1}}
\newcommand{\phalf}[1]{\cellcolor{orange!22}{\small #1}}
\newcommand{\pmost}[1]{\cellcolor{yellow!35}{\small #1}}
\newcommand{\pall}[1]{\cellcolor{green!22}{\small #1}}

\begin{table*}[t]
\centering
\caption{Pass rates of evaluated configurations on \nitr{} (21 cases).
\emph{Agent} uses agentic scaffolding; \emph{API} uses direct inference.
Per-dimension cell background: \colorbox{green!22}{\phantom{x}}~all pass,
\colorbox{yellow!35}{\phantom{x}}~majority,
\colorbox{orange!22}{\phantom{x}}~minority,
\colorbox{red!18}{\phantom{x}}~none.}
\label{tab:results}
\setlength{\tabcolsep}{4.5pt}
\begin{tabular*}{\textwidth}{@{\extracolsep{\fill}}llcccccccccccc}
\toprule
\textbf{Model} & \textbf{Mode} & \textbf{\#} & \textbf{Rate}
  & \textbf{D1} & \textbf{D2} & \textbf{D3}
  & \textbf{D4} & \textbf{D5} & \textbf{D6}
  & \textbf{D7} & \textbf{D8} & \textbf{D9} \\
\midrule
GPT-5.3-Cx       & Agent & 12 & 57\% & \pmost{2/3} & \phalf{1/3} & \phalf{1/2} & \phalf{1/2} & \phalf{1/2} & \phalf{1/2} & \pmost{2/3} & \pall{2/2}  & \phalf{1/2} \\
GPT-5.2-Cx       & Agent & 11 & 52\% & \pall{3/3}  & \phalf{1/3} & \pnone{0/2} & \phalf{1/2} & \phalf{1/2} & \phalf{1/2} & \pmost{2/3} & \phalf{1/2} & \phalf{1/2} \\
GPT-5            & Agent & 11 & 52\% & \pmost{2/3} & \phalf{1/3} & \phalf{1/2} & \phalf{1/2} & \phalf{1/2} & \pnone{0/2} & \pall{3/3}  & \phalf{1/2} & \phalf{1/2} \\
GPT-5.4          & Agent & 10 & 48\% & \pmost{2/3} & \phalf{1/3} & \phalf{1/2} & \phalf{1/2} & \phalf{1/2} & \pnone{0/2} & \pmost{2/3} & \phalf{1/2} & \phalf{1/2} \\
GPT-5.4          & API   &  7 & 33\% & \phalf{1/3} & \phalf{1/3} & \pnone{0/2} & \phalf{1/2} & \pnone{0/2} & \pnone{0/2} & \phalf{1/3} & \pall{2/2}  & \phalf{1/2} \\
GPT-5-Mini       & API   &  6 & 29\% & \pnone{0/3} & \phalf{1/3} & \pnone{0/2} & \phalf{1/2} & \phalf{1/2} & \pnone{0/2} & \phalf{1/3} & \phalf{1/2} & \phalf{1/2} \\
GPT-5.3-Cx       & API   &  6 & 29\% & \pnone{0/3} & \phalf{1/3} & \pnone{0/2} & \phalf{1/2} & \phalf{1/2} & \pnone{0/2} & \phalf{1/3} & \phalf{1/2} & \phalf{1/2} \\
\midrule
Claude Opus 4.6   & Agent & 12 & 57\% & \pall{3/3}  & \phalf{1/3} & \phalf{1/2} & \phalf{1/2} & \pall{2/2}  & \pnone{0/2} & \pmost{2/3} & \phalf{1/2} & \phalf{1/2} \\
Claude Opus 4.5   & Agent & 10 & 48\% & \pmost{2/3} & \phalf{1/3} & \phalf{1/2} & \phalf{1/2} & \phalf{1/2} & \pnone{0/2} & \pmost{2/3} & \phalf{1/2} & \phalf{1/2} \\
Claude Sonnet 4.6 & Agent &  9 & 43\% & \pmost{2/3} & \phalf{1/3} & \pnone{0/2} & \phalf{1/2} & \phalf{1/2} & \pnone{0/2} & \pmost{2/3} & \phalf{1/2} & \phalf{1/2} \\
Claude Sonnet 4.5 & Agent &  8 & 38\% & \pmost{2/3} & \phalf{1/3} & \phalf{1/2} & \pnone{0/2} & \phalf{1/2} & \pnone{0/2} & \pmost{2/3} & \phalf{1/2} & \pnone{0/2} \\
Claude Opus 4.5   & API   &  6 & 29\% & \phalf{1/3} & \phalf{1/3} & \pnone{0/2} & \phalf{1/2} & \pnone{0/2} & \pnone{0/2} & \phalf{1/3} & \phalf{1/2} & \phalf{1/2} \\
Claude Opus 4.6   & API   &  6 & 29\% & \phalf{1/3} & \phalf{1/3} & \pnone{0/2} & \phalf{1/2} & \pnone{0/2} & \pnone{0/2} & \phalf{1/3} & \phalf{1/2} & \phalf{1/2} \\
Claude Sonnet 4.5 & API   &  6 & 29\% & \phalf{1/3} & \phalf{1/3} & \pnone{0/2} & \phalf{1/2} & \pnone{0/2} & \pnone{0/2} & \phalf{1/3} & \phalf{1/2} & \phalf{1/2} \\
Claude Sonnet 4.6 & API   &  5 & 24\% & \pnone{0/3} & \phalf{1/3} & \pnone{0/2} & \phalf{1/2} & \pnone{0/2} & \pnone{0/2} & \phalf{1/3} & \phalf{1/2} & \phalf{1/2} \\
\midrule
Gemini 3.1 Pro    & Agent & 10 & 48\% & \pall{3/3}  & \pnone{0/3} & \pnone{0/2} & \phalf{1/2} & \phalf{1/2} & \pnone{0/2} & \phalf{1/3} & \pall{2/2}  & \pall{2/2}  \\
Gemini 3.1 Flash  & Agent &  6 & 29\% & \phalf{1/3} & \phalf{1/3} & \pnone{0/2} & \phalf{1/2} & \pnone{0/2} & \pnone{0/2} & \phalf{1/3} & \phalf{1/2} & \phalf{1/2} \\
Gemini 2.5 Pro    & Agent &  5 & 24\% & \phalf{1/3} & \phalf{1/3} & \pnone{0/2} & \phalf{1/2} & \pnone{0/2} & \pnone{0/2} & \phalf{1/3} & \phalf{1/2} & \pnone{0/2} \\
Gemini 3.1 Pro    & API   &  8 & 38\% & \pmost{2/3} & \phalf{1/3} & \phalf{1/2} & \phalf{1/2} & \pnone{0/2} & \pnone{0/2} & \phalf{1/3} & \phalf{1/2} & \phalf{1/2} \\
Gemini 2.5 Pro    & API   &  6 & 29\% & \pmost{2/3} & \phalf{1/3} & \pnone{0/2} & \pnone{0/2} & \pnone{0/2} & \pnone{0/2} & \pnone{0/3} & \pall{2/2}  & \phalf{1/2} \\
Gemini 3.1 Flash  & API   &  5 & 24\% & \phalf{1/3} & \phalf{1/3} & \pnone{0/2} & \pnone{0/2} & \pnone{0/2} & \pnone{0/2} & \phalf{1/3} & \phalf{1/2} & \phalf{1/2} \\
\midrule
Qwen3-Coder      & API   &  5 & 24\% & \phalf{1/3} & \phalf{1/3} & \pnone{0/2} & \pnone{0/2} & \pnone{0/2} & \pnone{0/2} & \phalf{1/3} & \phalf{1/2} & \phalf{1/2} \\
Qwen3-80B        & API   &  5 & 24\% & \pnone{0/3} & \phalf{1/3} & \pnone{0/2} & \phalf{1/2} & \pnone{0/2} & \pnone{0/2} & \phalf{1/3} & \phalf{1/2} & \phalf{1/2} \\
\midrule
\textit{Agent avg.}     & & \textit{9.5} & \textit{45\%} & & & & & & & & & \\
\textit{API avg.}       & & \textit{5.9} & \textit{28\%} & & & & & & & & & \\
\midrule
\textit{Case pass rate} & & & & \textit{48\%} & \textit{32\%} & \textit{15\%} & \textit{41\%} & \textit{26\%} & \textit{4\%} & \textit{45\%} & \textit{59\%} & \textit{48\%} \\
\bottomrule
\end{tabular*}
\end{table*}

\subsection{RQ1: Overall Performance}

Table~\ref{tab:results} summarizes performance across the 21-case suite.
Overall pass rates range from 24\% (5/21 cases) for the weakest configurations
to 57\% (12/21) for the strongest ones.
The best-performing systems are \texttt{GPT-5.3-Cx (Agt)} and \texttt{Claude Opus 4.6 (Agt)},
which each solve 12 of 21 cases.
The strongest API-only configuration, \texttt{Gemini 3.1 Pro (API)}, solves 8 of 21 cases (38\%).
Taken together, these results indicate that even frontier coding systems solve only about half of the
\textsc{NITR} suite at best.
Maintainability-preserving repository evolution therefore remains a substantial open challenge rather than a nearly solved capability.

Performance is highly uneven across the suite, suggesting that current systems succeed mainly when the repository already exposes a clear path for change, but struggle once they must infer and preserve that path themselves.
A small number of probes are close to saturation.
Case~012 (\emph{cache lifecycle}) is solved by all 23 configurations, while
Case~003 (\emph{reuse existing code}), Case~009 (\emph{session-expiry testability}),
and Case~010 (\emph{logging side-effects}) are passed by 21--22 systems.
These are cases where the starter repository already provides a strong structural path toward the desired change,
so the main burden is recognizing and following that path correctly.

At the other extreme, five probes are not solved by any configuration:
Case~002 (\emph{refactor-and-reuse}),
Case~004 (\emph{responsibility decomposition in the CV pipeline}),
Case~005 (\emph{pricing extension structure}),
Case~015 (\emph{pipeline provider decoupling}), and
Case~021 (\emph{inline filter entrypoint reuse}).
These universally unsolved cases are revealing because they are not simply hard programming tasks.
Rather, they require the model to preserve or recover an architectural direction under structural pressure:
refactor toward existing abstractions, maintain responsibility boundaries, extend behavior without patching central logic,
or route changes through the repository's intended reuse path.
This suggests that the main limitation exposed by \textsc{NITR} is not raw implementation ability alone,
but difficulty maintaining design discipline during repository-level change.

A further pattern emerges when we separate micro cases from multi-step cases, also shown in Figure~\ref{fig:granularity_passrate}.
Across all 23 configurations, micro cases are passed at 53.5\% (123/230), whereas
multi-step cases are passed at only 20.6\% (52/253).
The same drop appears in both evaluation regimes.
API-only systems fall from 44.2\% on micro cases (53/120) to 13.6\% on multi-step cases (18/132),
while agent-mode systems fall from 63.6\% (70/110) to 28.1\% (34/121).
Moreover, four of the five universally unsolved cases belong to the multi-step setting.

\begin{tcolorbox}[
  floatplacement=tb,
  colback=gray!6,
  colframe=black!70,
  boxrule=0.6pt,
  arc=2pt,
  left=6pt,
  right=6pt,
  top=6pt,
  bottom=6pt,
  title={Summary 1},
  fonttitle=\bfseries,
]
\textsc{NITR} shows that current AI coding tools remain far from robust:
they solve only 36.2\% of cases, and pass rates collapse from 53.5\% on micro cases to 20.6\% on multi-step cases.
The real bottleneck is sustained structural discipline under change, not isolated code generation or bug fixing.
\end{tcolorbox}

\subsection{RQ2: Which Maintainability Dimensions Are Most Challenging?}
\vspace{.3em}
Table~\ref{tab:results} (bottom row) reports aggregate pass rates by maintainability dimension. The results reveal a non-uniform difficulty pattern within NITR rather than a uniform decline across the suite. Please note that because some dimensions are represented by a few cases, these rates should be read as diagnostic trends within {\sc NITR} rather than high-confidence population estimates of maintainability difficulty.

\emph{Dependency Control} (D6) is the hardest dimension by a wide margin, with a pass rate of just 4.3\%.
Across its two cases (Case~008 and Case~015), only 2 of 46 evaluated attempts succeed.
\emph{Responsibility Decomposition} (D3) is the second hardest at 15.2\%, followed by
\emph{Interface and Substitutability Discipline} (D5) at 26.1\%, and
\emph{Reuse and Repo Awareness} (D2) at 31.9\%.
These are dimensions that require the model to identify and preserve latent repository structure:
respect dependency boundaries, maintain responsibility separation, extend behavior through the right abstraction,
and reuse existing mechanisms instead of introducing parallel paths.

By contrast, the easiest dimension is \emph{State Ownership and Lifecycle} (D8) at 58.7\%.
It is followed by \emph{Change Locality} (D1) and \emph{Side-Effect Isolation} (D9), both at 47.8\%, and
\emph{Testability and Determinism} (D7) at 44.9\%.
These dimensions are still nontrivial, but in many cases the intended structural direction is more explicit in the starter repository.

\begin{tcolorbox}[
  floatplacement=tb,
  colback=gray!6,
  colframe=black!70,
  boxrule=0.6pt,
  arc=2pt,
  left=6pt,
  right=6pt,
  top=6pt,
  bottom=6pt,
  title={Summary 2},
  fonttitle=\bfseries,
]
Within {\sc NITR}, maintainability difficulty is highly concentrated instead of evenly distributed. Current AI coding tools achieve 58.7\% on D8 but only 4.3\% on D6, a 13$\times$ gap.
Within NITR, the bottlenecks are dependency control, responsibility decomposition, and abstraction-respecting extension.\end{tcolorbox}

\begin{figure*}[t]
\centering
\caption{Pass/fail heatmap of 23 evaluated configurations across 21 cases.
\colorbox{passcolor!60}{\phantom{X}}~Pass,
~\colorbox{structcolor!60}{\phantom{X}}~S-category failure (functional tests pass, maintainability oracle fails),
~\colorbox{funccolor!60}{\phantom{X}}~Functional test failure,
~\colorbox{bothcolor!60}{\phantom{X}}~Both fail,
~\colorbox{buildcolor!60}{\phantom{X}}~Build failure.
Cases marked \colorbox{structcolor!60}{\phantom{X}} confirm that functional tests alone
miss these maintainability failures, which are detectable only via structural probes.}
\label{fig:heatmap}
\vspace{.5em}
\resizebox{\textwidth}{!}{%
\setlength{\tabcolsep}{3pt}
\fontsize{7pt}{9pt}\selectfont
\begin{tabular}{r|ccccccccccccccccccccc}
\toprule
\textbf{Model} & \rotatebox{90}{\textbf{001}} & \rotatebox{90}{\textbf{002}} & \rotatebox{90}{\textbf{003}} & \rotatebox{90}{\textbf{004}} & \rotatebox{90}{\textbf{005}} & \rotatebox{90}{\textbf{006}} & \rotatebox{90}{\textbf{007}} & \rotatebox{90}{\textbf{008}} & \rotatebox{90}{\textbf{009}} & \rotatebox{90}{\textbf{010}} & \rotatebox{90}{\textbf{011}} & \rotatebox{90}{\textbf{012}} & \rotatebox{90}{\textbf{013}} & \rotatebox{90}{\textbf{014}} & \rotatebox{90}{\textbf{015}} & \rotatebox{90}{\textbf{016}} & \rotatebox{90}{\textbf{017}} & \rotatebox{90}{\textbf{018}} & \rotatebox{90}{\textbf{019}} & \rotatebox{90}{\textbf{020}} & \rotatebox{90}{\textbf{021}} \\
\midrule
\texttt{GPT-5.3-Cx (Agt)} & \cellcolor{structcolor!60}S & \cellcolor{structcolor!60}S & \cellcolor{passcolor!60}\ding{51} & \cellcolor{funccolor!60}F & \cellcolor{bothcolor!60}B & \cellcolor{structcolor!60}S & \cellcolor{passcolor!60}\ding{51} & \cellcolor{passcolor!60}\ding{51} & \cellcolor{passcolor!60}\ding{51} & \cellcolor{passcolor!60}\ding{51} & \cellcolor{passcolor!60}\ding{51} & \cellcolor{passcolor!60}\ding{51} & \cellcolor{passcolor!60}\ding{51} & \cellcolor{passcolor!60}\ding{51} & \cellcolor{funccolor!60}F & \cellcolor{passcolor!60}\ding{51} & \cellcolor{passcolor!60}\ding{51} & \cellcolor{funccolor!60}F & \cellcolor{funccolor!60}F & \cellcolor{passcolor!60}\ding{51} & \cellcolor{funccolor!60}F \\
\texttt{GPT-5.3-Cx (API)} & \cellcolor{structcolor!60}S & \cellcolor{structcolor!60}S & \cellcolor{passcolor!60}\ding{51} & \cellcolor{funccolor!60}F & \cellcolor{bothcolor!60}B & \cellcolor{bothcolor!60}B & \cellcolor{passcolor!60}\ding{51} & \cellcolor{funccolor!60}F & \cellcolor{passcolor!60}\ding{51} & \cellcolor{passcolor!60}\ding{51} & \cellcolor{funccolor!60}F & \cellcolor{passcolor!60}\ding{51} & \cellcolor{funccolor!60}F & \cellcolor{passcolor!60}\ding{51} & \cellcolor{funccolor!60}F & \cellcolor{funccolor!60}F & \cellcolor{structcolor!60}S & \cellcolor{funccolor!60}F & \cellcolor{bothcolor!60}B & \cellcolor{structcolor!60}S & \cellcolor{funccolor!60}F \\
\texttt{GPT-5.4 (Agt)} & \cellcolor{structcolor!60}S & \cellcolor{structcolor!60}S & \cellcolor{passcolor!60}\ding{51} & \cellcolor{funccolor!60}F & \cellcolor{bothcolor!60}B & \cellcolor{structcolor!60}S & \cellcolor{passcolor!60}\ding{51} & \cellcolor{funccolor!60}F & \cellcolor{passcolor!60}\ding{51} & \cellcolor{passcolor!60}\ding{51} & \cellcolor{passcolor!60}\ding{51} & \cellcolor{passcolor!60}\ding{51} & \cellcolor{passcolor!60}\ding{51} & \cellcolor{passcolor!60}\ding{51} & \cellcolor{funccolor!60}F & \cellcolor{passcolor!60}\ding{51} & \cellcolor{structcolor!60}S & \cellcolor{funccolor!60}F & \cellcolor{bothcolor!60}B & \cellcolor{passcolor!60}\ding{51} & \cellcolor{funccolor!60}F \\
\texttt{GPT-5.4 (API)} & \cellcolor{structcolor!60}S & \cellcolor{structcolor!60}S & \cellcolor{passcolor!60}\ding{51} & \cellcolor{funccolor!60}F & \cellcolor{bothcolor!60}B & \cellcolor{buildcolor!60}E & \cellcolor{funccolor!60}F & \cellcolor{funccolor!60}F & \cellcolor{passcolor!60}\ding{51} & \cellcolor{passcolor!60}\ding{51} & \cellcolor{funccolor!60}F & \cellcolor{passcolor!60}\ding{51} & \cellcolor{passcolor!60}\ding{51} & \cellcolor{passcolor!60}\ding{51} & \cellcolor{funccolor!60}F & \cellcolor{funccolor!60}F & \cellcolor{passcolor!60}\ding{51} & \cellcolor{funccolor!60}F & \cellcolor{bothcolor!60}B & \cellcolor{structcolor!60}S & \cellcolor{funccolor!60}F \\
\texttt{GPT-5.2-Cx (Agt)} & \cellcolor{passcolor!60}\ding{51} & \cellcolor{bothcolor!60}B & \cellcolor{passcolor!60}\ding{51} & \cellcolor{funccolor!60}F & \cellcolor{bothcolor!60}B & \cellcolor{structcolor!60}S & \cellcolor{passcolor!60}\ding{51} & \cellcolor{passcolor!60}\ding{51} & \cellcolor{passcolor!60}\ding{51} & \cellcolor{passcolor!60}\ding{51} & \cellcolor{passcolor!60}\ding{51} & \cellcolor{passcolor!60}\ding{51} & \cellcolor{passcolor!60}\ding{51} & \cellcolor{passcolor!60}\ding{51} & \cellcolor{funccolor!60}F & \cellcolor{passcolor!60}\ding{51} & \cellcolor{structcolor!60}S & \cellcolor{funccolor!60}F & \cellcolor{bothcolor!60}B & \cellcolor{structcolor!60}S & \cellcolor{funccolor!60}F \\
\texttt{GPT-5 (Agt)} & \cellcolor{structcolor!60}S & \cellcolor{bothcolor!60}B & \cellcolor{passcolor!60}\ding{51} & \cellcolor{buildcolor!60}E & \cellcolor{bothcolor!60}B & \cellcolor{structcolor!60}S & \cellcolor{passcolor!60}\ding{51} & \cellcolor{bothcolor!60}B & \cellcolor{passcolor!60}\ding{51} & \cellcolor{passcolor!60}\ding{51} & \cellcolor{passcolor!60}\ding{51} & \cellcolor{passcolor!60}\ding{51} & \cellcolor{passcolor!60}\ding{51} & \cellcolor{passcolor!60}\ding{51} & \cellcolor{funccolor!60}F & \cellcolor{passcolor!60}\ding{51} & \cellcolor{structcolor!60}S & \cellcolor{passcolor!60}\ding{51} & \cellcolor{bothcolor!60}B & \cellcolor{passcolor!60}\ding{51} & \cellcolor{funccolor!60}F \\
\texttt{GPT-5-Mini (API)} & \cellcolor{structcolor!60}S & \cellcolor{bothcolor!60}B & \cellcolor{passcolor!60}\ding{51} & \cellcolor{funccolor!60}F & \cellcolor{buildcolor!60}E & \cellcolor{buildcolor!60}E & \cellcolor{passcolor!60}\ding{51} & \cellcolor{bothcolor!60}B & \cellcolor{passcolor!60}\ding{51} & \cellcolor{passcolor!60}\ding{51} & \cellcolor{funccolor!60}F & \cellcolor{passcolor!60}\ding{51} & \cellcolor{funccolor!60}F & \cellcolor{passcolor!60}\ding{51} & \cellcolor{funccolor!60}F & \cellcolor{funccolor!60}F & \cellcolor{funccolor!60}F & \cellcolor{funccolor!60}F & \cellcolor{bothcolor!60}B & \cellcolor{structcolor!60}S & \cellcolor{funccolor!60}F \\
\texttt{Claude Opus 4.5 (Agt)} & \cellcolor{structcolor!60}S & \cellcolor{bothcolor!60}B & \cellcolor{passcolor!60}\ding{51} & \cellcolor{funccolor!60}F & \cellcolor{funccolor!60}F & \cellcolor{structcolor!60}S & \cellcolor{passcolor!60}\ding{51} & \cellcolor{funccolor!60}F & \cellcolor{passcolor!60}\ding{51} & \cellcolor{passcolor!60}\ding{51} & \cellcolor{passcolor!60}\ding{51} & \cellcolor{passcolor!60}\ding{51} & \cellcolor{passcolor!60}\ding{51} & \cellcolor{passcolor!60}\ding{51} & \cellcolor{funccolor!60}F & \cellcolor{passcolor!60}\ding{51} & \cellcolor{structcolor!60}S & \cellcolor{funccolor!60}F & \cellcolor{bothcolor!60}B & \cellcolor{passcolor!60}\ding{51} & \cellcolor{funccolor!60}F \\
\texttt{Claude Opus 4.5 (API)} & \cellcolor{structcolor!60}S & \cellcolor{structcolor!60}S & \cellcolor{passcolor!60}\ding{51} & \cellcolor{funccolor!60}F & \cellcolor{funccolor!60}F & \cellcolor{bothcolor!60}B & \cellcolor{funccolor!60}F & \cellcolor{bothcolor!60}B & \cellcolor{passcolor!60}\ding{51} & \cellcolor{passcolor!60}\ding{51} & \cellcolor{funccolor!60}F & \cellcolor{passcolor!60}\ding{51} & \cellcolor{passcolor!60}\ding{51} & \cellcolor{passcolor!60}\ding{51} & \cellcolor{funccolor!60}F & \cellcolor{funccolor!60}F & \cellcolor{bothcolor!60}B & \cellcolor{funccolor!60}F & \cellcolor{bothcolor!60}B & \cellcolor{structcolor!60}S & \cellcolor{funccolor!60}F \\
\texttt{Claude Opus 4.6 (Agt)} & \cellcolor{passcolor!60}\ding{51} & \cellcolor{structcolor!60}S & \cellcolor{passcolor!60}\ding{51} & \cellcolor{funccolor!60}F & \cellcolor{bothcolor!60}B & \cellcolor{passcolor!60}\ding{51} & \cellcolor{passcolor!60}\ding{51} & \cellcolor{funccolor!60}F & \cellcolor{passcolor!60}\ding{51} & \cellcolor{passcolor!60}\ding{51} & \cellcolor{passcolor!60}\ding{51} & \cellcolor{passcolor!60}\ding{51} & \cellcolor{passcolor!60}\ding{51} & \cellcolor{passcolor!60}\ding{51} & \cellcolor{funccolor!60}F & \cellcolor{passcolor!60}\ding{51} & \cellcolor{bothcolor!60}B & \cellcolor{funccolor!60}F & \cellcolor{bothcolor!60}B & \cellcolor{passcolor!60}\ding{51} & \cellcolor{funccolor!60}F \\
\texttt{Claude Opus 4.6 (API)} & \cellcolor{structcolor!60}S & \cellcolor{structcolor!60}S & \cellcolor{passcolor!60}\ding{51} & \cellcolor{buildcolor!60}E & \cellcolor{funccolor!60}F & \cellcolor{bothcolor!60}B & \cellcolor{funccolor!60}F & \cellcolor{funccolor!60}F & \cellcolor{passcolor!60}\ding{51} & \cellcolor{passcolor!60}\ding{51} & \cellcolor{funccolor!60}F & \cellcolor{passcolor!60}\ding{51} & \cellcolor{passcolor!60}\ding{51} & \cellcolor{passcolor!60}\ding{51} & \cellcolor{funccolor!60}F & \cellcolor{funccolor!60}F & \cellcolor{bothcolor!60}B & \cellcolor{funccolor!60}F & \cellcolor{bothcolor!60}B & \cellcolor{structcolor!60}S & \cellcolor{funccolor!60}F \\
\texttt{Claude Sonnet 4.5 (Agt)} & \cellcolor{structcolor!60}S & \cellcolor{structcolor!60}S & \cellcolor{passcolor!60}\ding{51} & \cellcolor{funccolor!60}F & \cellcolor{bothcolor!60}B & \cellcolor{buildcolor!60}E & \cellcolor{passcolor!60}\ding{51} & \cellcolor{buildcolor!60}E & \cellcolor{passcolor!60}\ding{51} & \cellcolor{funccolor!60}F & \cellcolor{passcolor!60}\ding{51} & \cellcolor{passcolor!60}\ding{51} & \cellcolor{passcolor!60}\ding{51} & \cellcolor{buildcolor!60}E & \cellcolor{funccolor!60}F & \cellcolor{passcolor!60}\ding{51} & \cellcolor{bothcolor!60}B & \cellcolor{funccolor!60}F & \cellcolor{bothcolor!60}B & \cellcolor{passcolor!60}\ding{51} & \cellcolor{funccolor!60}F \\
\texttt{Claude Sonnet 4.5 (API)} & \cellcolor{structcolor!60}S & \cellcolor{structcolor!60}S & \cellcolor{passcolor!60}\ding{51} & \cellcolor{funccolor!60}F & \cellcolor{buildcolor!60}E & \cellcolor{funccolor!60}F & \cellcolor{bothcolor!60}B & \cellcolor{funccolor!60}F & \cellcolor{passcolor!60}\ding{51} & \cellcolor{passcolor!60}\ding{51} & \cellcolor{funccolor!60}F & \cellcolor{passcolor!60}\ding{51} & \cellcolor{passcolor!60}\ding{51} & \cellcolor{passcolor!60}\ding{51} & \cellcolor{funccolor!60}F & \cellcolor{funccolor!60}F & \cellcolor{bothcolor!60}B & \cellcolor{funccolor!60}F & \cellcolor{bothcolor!60}B & \cellcolor{structcolor!60}S & \cellcolor{funccolor!60}F \\
\texttt{Claude Sonnet 4.6 (Agt)} & \cellcolor{structcolor!60}S & \cellcolor{structcolor!60}S & \cellcolor{passcolor!60}\ding{51} & \cellcolor{funccolor!60}F & \cellcolor{funccolor!60}F & \cellcolor{passcolor!60}\ding{51} & \cellcolor{funccolor!60}F & \cellcolor{funccolor!60}F & \cellcolor{passcolor!60}\ding{51} & \cellcolor{passcolor!60}\ding{51} & \cellcolor{passcolor!60}\ding{51} & \cellcolor{passcolor!60}\ding{51} & \cellcolor{passcolor!60}\ding{51} & \cellcolor{passcolor!60}\ding{51} & \cellcolor{funccolor!60}F & \cellcolor{passcolor!60}\ding{51} & \cellcolor{bothcolor!60}B & \cellcolor{funccolor!60}F & \cellcolor{bothcolor!60}B & \cellcolor{structcolor!60}S & \cellcolor{funccolor!60}F \\
\texttt{Claude Sonnet 4.6 (API)} & \cellcolor{structcolor!60}S & \cellcolor{structcolor!60}S & \cellcolor{passcolor!60}\ding{51} & \cellcolor{buildcolor!60}E & \cellcolor{buildcolor!60}E & \cellcolor{funccolor!60}F & \cellcolor{bothcolor!60}B & \cellcolor{funccolor!60}F & \cellcolor{passcolor!60}\ding{51} & \cellcolor{passcolor!60}\ding{51} & \cellcolor{funccolor!60}F & \cellcolor{passcolor!60}\ding{51} & \cellcolor{funccolor!60}F & \cellcolor{passcolor!60}\ding{51} & \cellcolor{funccolor!60}F & \cellcolor{funccolor!60}F & \cellcolor{structcolor!60}S & \cellcolor{funccolor!60}F & \cellcolor{bothcolor!60}B & \cellcolor{structcolor!60}S & \cellcolor{funccolor!60}F \\
\texttt{Gemini 3.1 Pro (Agt)} & \cellcolor{passcolor!60}\ding{51} & \cellcolor{buildcolor!60}E & \cellcolor{funccolor!60}F & \cellcolor{funccolor!60}F & \cellcolor{funccolor!60}F & \cellcolor{structcolor!60}S & \cellcolor{passcolor!60}\ding{51} & \cellcolor{structcolor!60}S & \cellcolor{funccolor!60}F & \cellcolor{passcolor!60}\ding{51} & \cellcolor{passcolor!60}\ding{51} & \cellcolor{passcolor!60}\ding{51} & \cellcolor{passcolor!60}\ding{51} & \cellcolor{passcolor!60}\ding{51} & \cellcolor{structcolor!60}S & \cellcolor{bothcolor!60}B & \cellcolor{passcolor!60}\ding{51} & \cellcolor{passcolor!60}\ding{51} & \cellcolor{passcolor!60}\ding{51} & \cellcolor{structcolor!60}S & \cellcolor{funccolor!60}F \\
\texttt{Gemini 3.1 Pro (API)} & \cellcolor{structcolor!60}S & \cellcolor{structcolor!60}S & \cellcolor{passcolor!60}\ding{51} & \cellcolor{funccolor!60}F & \cellcolor{funccolor!60}F & \cellcolor{buildcolor!60}E & \cellcolor{funccolor!60}F & \cellcolor{structcolor!60}S & \cellcolor{passcolor!60}\ding{51} & \cellcolor{passcolor!60}\ding{51} & \cellcolor{passcolor!60}\ding{51} & \cellcolor{passcolor!60}\ding{51} & \cellcolor{passcolor!60}\ding{51} & \cellcolor{passcolor!60}\ding{51} & \cellcolor{funccolor!60}F & \cellcolor{funccolor!60}F & \cellcolor{structcolor!60}S & \cellcolor{funccolor!60}F & \cellcolor{bothcolor!60}B & \cellcolor{passcolor!60}\ding{51} & \cellcolor{funccolor!60}F \\
\texttt{Gemini 3.1 Flash (Agt)} & \cellcolor{structcolor!60}S & \cellcolor{structcolor!60}S & \cellcolor{passcolor!60}\ding{51} & \cellcolor{funccolor!60}F & \cellcolor{funccolor!60}F & \cellcolor{funccolor!60}F & \cellcolor{funccolor!60}F & \cellcolor{funccolor!60}F & \cellcolor{passcolor!60}\ding{51} & \cellcolor{passcolor!60}\ding{51} & \cellcolor{funccolor!60}F & \cellcolor{passcolor!60}\ding{51} & \cellcolor{passcolor!60}\ding{51} & \cellcolor{passcolor!60}\ding{51} & \cellcolor{funccolor!60}F & \cellcolor{funccolor!60}F & \cellcolor{bothcolor!60}B & \cellcolor{funccolor!60}F & \cellcolor{buildcolor!60}E & \cellcolor{structcolor!60}S & \cellcolor{funccolor!60}F \\
\texttt{Gemini 3.1 Flash (API)} & \cellcolor{structcolor!60}S & \cellcolor{structcolor!60}S & \cellcolor{passcolor!60}\ding{51} & \cellcolor{buildcolor!60}E & \cellcolor{funccolor!60}F & \cellcolor{buildcolor!60}E & \cellcolor{funccolor!60}F & \cellcolor{funccolor!60}F & \cellcolor{passcolor!60}\ding{51} & \cellcolor{passcolor!60}\ding{51} & \cellcolor{funccolor!60}F & \cellcolor{passcolor!60}\ding{51} & \cellcolor{passcolor!60}\ding{51} & \cellcolor{funccolor!60}F & \cellcolor{funccolor!60}F & \cellcolor{funccolor!60}F & \cellcolor{bothcolor!60}B & \cellcolor{funccolor!60}F & \cellcolor{buildcolor!60}E & \cellcolor{structcolor!60}S & \cellcolor{funccolor!60}F \\
\texttt{Gemini 2.5 Pro (Agt)} & \cellcolor{bothcolor!60}B & \cellcolor{buildcolor!60}E & \cellcolor{passcolor!60}\ding{51} & \cellcolor{buildcolor!60}E & \cellcolor{buildcolor!60}E & \cellcolor{buildcolor!60}E & \cellcolor{funccolor!60}F & \cellcolor{funccolor!60}F & \cellcolor{passcolor!60}\ding{51} & \cellcolor{funccolor!60}F & \cellcolor{funccolor!60}F & \cellcolor{passcolor!60}\ding{51} & \cellcolor{passcolor!60}\ding{51} & \cellcolor{passcolor!60}\ding{51} & \cellcolor{funccolor!60}F & \cellcolor{buildcolor!60}E & \cellcolor{buildcolor!60}E & \cellcolor{funccolor!60}F & \cellcolor{buildcolor!60}E & \cellcolor{structcolor!60}S & \cellcolor{funccolor!60}F \\
\texttt{Gemini 2.5 Pro (API)} & \cellcolor{passcolor!60}\ding{51} & \cellcolor{structcolor!60}S & \cellcolor{passcolor!60}\ding{51} & \cellcolor{funccolor!60}F & \cellcolor{buildcolor!60}E & \cellcolor{bothcolor!60}B & \cellcolor{buildcolor!60}E & \cellcolor{funccolor!60}F & \cellcolor{buildcolor!60}E & \cellcolor{passcolor!60}\ding{51} & \cellcolor{funccolor!60}F & \cellcolor{passcolor!60}\ding{51} & \cellcolor{passcolor!60}\ding{51} & \cellcolor{funccolor!60}F & \cellcolor{funccolor!60}F & \cellcolor{buildcolor!60}E & \cellcolor{passcolor!60}\ding{51} & \cellcolor{funccolor!60}F & \cellcolor{buildcolor!60}E & \cellcolor{structcolor!60}S & \cellcolor{funccolor!60}F \\
\texttt{Qwen3-Coder (API)} & \cellcolor{passcolor!60}\ding{51} & \cellcolor{buildcolor!60}E & \cellcolor{passcolor!60}\ding{51} & \cellcolor{buildcolor!60}E & \cellcolor{bothcolor!60}B & \cellcolor{buildcolor!60}E & \cellcolor{funccolor!60}F & \cellcolor{buildcolor!60}E & \cellcolor{passcolor!60}\ding{51} & \cellcolor{passcolor!60}\ding{51} & \cellcolor{funccolor!60}F & \cellcolor{passcolor!60}\ding{51} & \cellcolor{funccolor!60}F & \cellcolor{buildcolor!60}E & \cellcolor{funccolor!60}F & \cellcolor{funccolor!60}F & \cellcolor{buildcolor!60}E & \cellcolor{funccolor!60}F & \cellcolor{bothcolor!60}B & \cellcolor{buildcolor!60}E & \cellcolor{funccolor!60}F \\
\texttt{Qwen3-80B (API)} & \cellcolor{structcolor!60}S & \cellcolor{structcolor!60}S & \cellcolor{passcolor!60}\ding{51} & \cellcolor{buildcolor!60}E & \cellcolor{buildcolor!60}E & \cellcolor{buildcolor!60}E & \cellcolor{funccolor!60}F & \cellcolor{buildcolor!60}E & \cellcolor{passcolor!60}\ding{51} & \cellcolor{passcolor!60}\ding{51} & \cellcolor{funccolor!60}F & \cellcolor{passcolor!60}\ding{51} & \cellcolor{funccolor!60}F & \cellcolor{passcolor!60}\ding{51} & \cellcolor{buildcolor!60}E & \cellcolor{buildcolor!60}E & \cellcolor{bothcolor!60}B & \cellcolor{funccolor!60}F & \cellcolor{bothcolor!60}B & \cellcolor{structcolor!60}S & \cellcolor{funccolor!60}F \\
\midrule
\textit{Pass} & 5/23 & 0/23 & 22/23 & 0/23 & 0/23 & 2/23 & 10/23 & 2/23 & 21/23 & 21/23 & 10/23 & 23/23 & 18/23 & 19/23 & 0/23 & 8/23 & 4/23 & 2/23 & 1/23 & 7/23 & 0/23 \\
\bottomrule
\end{tabular}%
}
\vspace{.3em}
\end{figure*}

\vspace{.3em}

\subsection{RQ3: Maintainability Failures Missed by Functional Correctness}
To answer RQ3, we analyze whether systems fail and \emph{how} they fail.
In particular, we distinguish ordinary functional failures from failures that remain behaviorally correct yet violate the maintainability constraint encoded by the probe.

Figure~\ref{fig:heatmap} summarizes the outcome of every (model, case) pair.
We classify each result as \textbf{Pass}, \textbf{F} (functional test failure),
\textbf{S} (functional tests pass, but the maintainability oracle fails),
\textbf{B} (both fail), or \textbf{E} (build error).
Across all 483 evaluated pairs, 36.2\% pass, 31.5\% are functional failures,
13.3\% are \textbf{S}-category failures, 10.4\% are joint failures, and 8.7\% are build errors. By this decoupling, {\sc NITR} is deliberately constructed so that a behaviorally correct shortcut and a maintainable repository-aligned solution can both satisfy the immediate task, while differing in long-term structural quality. 

The \textbf{S} category is especially important for answering this research question.
These are solutions that compile, pass the hidden functional tests, and would be counted as successful under conventional test-based evaluation, yet still violate the intended maintainability constraint.
Across the full suite, this occurs in 64 of 483 evaluated pairs (13.3\%).
The effect is particularly strong in Case~001, Case~002, and Case~020, which exhibit 17/23, 16/23, and 15/23 \textbf{S}-category outcomes, respectively.
In these probes, many systems produce code that appears correct behaviorally while silently introducing structural debt.
This shows that functional correctness and maintainability quality often decouple, and that test-only evaluation systematically overestimates maintainability competence on a nontrivial portion of the suite.

\vspace{.3em}
\noindent{\bf Manual Inspection.} To contextualize the failure matrix, we qualitatively inspected failed outcomes at the level of individual (model, case) runs. Across the 308 failed runs, the breakdowns clustered into a small set of recurring cross-case archetypes rather than appearing arbitrary or case-specific. These archetypes capture \emph{empirical failure mechanisms} shared across maintainability dimensions: reuse and dependency-control probes often fail through shortcut-over-reuse or incomplete abstraction refactors, whereas responsibility and side-effect probes more often fail through boundary contamination. Overall, \textsc{NITR} suggests that {\it current AI coding tools often preserve required behavior through structurally convenient shortcuts that accumulate maintainability debt beyond what functional tests alone expose.}

\begin{tcolorbox}[
  floatplacement=tb,
  colback=gray!6,
  colframe=black!70,
  boxrule=0.6pt,
  arc=2pt,
  left=6pt,
  right=6pt,
  top=6pt,
  bottom=6pt,
  title={Summary 3},
  fonttitle=\bfseries,
]
Within \textsc{NITR}, test passing is an unreliable proxy for maintainability preservation.
64/483 outcomes (13.3\%) are behaviorally correct yet structurally wrong, and 308 failed runs collapse into five recurring archetypes rather than case-specific noise.
\end{tcolorbox}

\begin{table*}[t]
\centering
\caption{The table summarizes recurring qualitative patterns and representative cases; it is not an exhaustive coded annotation of all failures.}
\label{tab:failure_archetypes}
\resizebox{\textwidth}{!}{
\begin{tabular}{p{4.5cm} p{6.9cm} p{3.2cm} p{1.5cm}}
\toprule
\textbf{Failure archetype} & \textbf{Defining behavior} & \textbf{Representative cases} & \textbf{Outcome} \\
\midrule
Shortcut over reuse or abstraction
& Clones logic or creates a parallel path instead of routing the change through a reusable or repository-aligned abstraction
& 001, 002, 011, 021
& S / F \\
\midrule
Boundary contamination
& Preserves visible behavior but violates responsibility, ownership, or side-effect boundaries
& 004, 019, 020
& S / F / B \\
\midrule
Incomplete abstraction refactor
& Starts the right abstraction shape but does not propagate the change consistently across the repository
& 006, 008, 014, 015
& E / B \\
\midrule
Later-step regression
& Works for earlier milestones but breaks baseline behavior or collapses under later extensions
& 005, 007, 016, 017, 018, 019
& F / B \\
\midrule
Missing test seam
& Leaves time, randomness, or configuration-sensitive behavior hardwired in core logic
& 009, 016, 018
& F \\
\bottomrule
\end{tabular}
}
\end{table*}

\subsection{RQ4: Agents vs.\ APIs Under the {\sc NITR} Harness}

Under our constrained harness, agent-mode configurations are associated with higher overall performance on \textsc{NITR}. We report this as an observational comparison rather than a clean causal estimate of scaffolding alone, since agent and API settings differ in interface and execution surface.
Across all configurations, agent mode averages 45.0\% (9.45/21 cases) versus 28.2\% (5.92/21) for API-only systems, a gain of 16.8 percentage points.
The best agent configuration solves 12 of 21 cases, compared with 8 for the best API-only configuration.

The same pattern appears in matched same-family comparisons.
Among the nine families evaluated in both modes, agent mode improves performance in eight, with an average gain of 3.0 passed cases.
For example, \texttt{GPT-5.3-Cx} and \texttt{Claude Opus 4.6} improve from 6 to 12 passed cases in agent mode.
Across {\sc NITR}, the largest gains appear on Cases~007, 011, 016, and 020, which require coordination across module boundaries, staged modifications, or responsi-\\bility-preserving changes.

However, agent mode does not remove the core bottlenecks exposed by {\sc NITR}: no matched family solves Cases~002, 004, 005, 015, or 021 in either mode.
Overall, under this harness, agent-mode systems are associated with better maintainability-preserving performance, but they do not resolve the underlying architectural failure modes.

\begin{tcolorbox}[
  floatplacement=tb,
  colback=gray!6,
  colframe=black!70,
  boxrule=0.6pt,
  arc=2pt,
  left=6pt,
  right=6pt,
  top=6pt,
  bottom=6pt,
  title={Summary 4},
  fonttitle=\bfseries,
]
Under the NITR harness, agent-mode configurations generally perform better than API-only ones (45.0\% vs. 28.2\%), but this does not isolate scaffolding as the sole cause and does not remove the core architectural bottlenecks. 
\end{tcolorbox}

\subsection{Representative Case Studies}
We discuss representative cases to illustrate distinct maintainability failure modes. Cases~001 and~020 capture {\sc NITR}'s central behavioral--structural gap, while Case~006 highlights brittle extension against an existing repository convention.

\vspace{.3em}
\noindent\textbf{Case 001 (D1, S-category maintainability failure).}
The task asks agents to implement a multi-type \codefontsmall{add} function
across three incremental steps.
Although the locked \codefontsmall{app/main.cc} calls \codefontsmall{add} with all
four types from step~1, only the integer path is tested initially, so explicit
overloads can still pass early.
By step~3, 17 of 23 configurations incur structural debt by introducing duplicated,
type-specific implementations while still passing the functional tests.
The structural probe requires a generic definition (\eg template or
\codefontsmall{auto}) and flags multiple explicit overloads as
\codefontsmall{ERR\_DUPLICATED\_IMPLEMENTATION}.

\vspace{.3em}
\noindent\textbf{Case 020 (D3, S-category maintainability failure).}
The task asks agents to implement a handover-packet ownership boundary between two
adjacent pipeline modules.
7 configurations pass the case.
15 of 23 configurations produce functionally correct solutions, but the
structural probe detects that the ownership-transfer function also performs
validation or enrichment that belongs in the receiving module, violating the
intended responsibility boundary.

\vspace{.3em}
\noindent\textbf{Case 006 (D5, mixed maintainability/functional/build failure).}
The task asks agents to refactor a hit-processing pipeline so that the sort stage
consumes a narrow \emph{view} type rather than the full \codefontsmall{HitBuffer}.
Two configurations pass.
6 configurations still pass all functional tests, but introduce no new view type.
Here the repository already establishes an extension contract through its existing
type-family structure, and the task requires the new sort-stage interface to follow
that pattern.
The structural probe therefore checks for a corresponding view abstraction in the
public interface using repository-consistent patterns (\eg \codefontsmall{SortView}
or \codefontsmall{IHitSorting}); this is not a stylistic naming preference, but a
check that the edit preserves the repository's extension structure rather than
taking an ad hoc shortcut.
A further four configurations introduce a view type but wire it incorrectly, breaking
functionality, and eight fail to compile altogether.

\section{Discussion and Implications}\label{sec:discussion}

\subsection{Implications}
NITR suggests implications for {\it education}, {\it code review}, and {\it agent design}. First, {\sc NITR} can be useful as a teaching instrument for SE courses. Its cases make design tradeoffs concrete by contrasting behaviorally correct solutions with structurally poor ones, which help students discuss reuse, boundaries, and testability in a more grounded way. Second, in practice, reviewing AI-generated code should go beyond asking whether a patch works, and ask whether it preserves reuse paths, dependency boundaries, responsibility separation, etc. Especially for multi-step changes, AI-generated code should be treated less as finished output and more as a structurally untrusted proposal. Third, the failures exposed by {\sc NITR} suggest that improving coding agents will require stronger repository-level structural reasoning rather than better local code synthesis alone.

\subsection{Threats to Validity}
\textbf{Construct validity.} \textsc{NITR} operationalizes maintainability through authored probes and case-specific structural oracles. This means the framework measures maintainability preservation as instantiated by explicit repository-evolution pressures and executable design-boundary checks, not maintainability in an unrestricted sense. We therefore position \textsc{NITR} as a diagnostic instrument for targeted maintainability pressures rather than a universal maintainability benchmark. \textbf{Internal validity.} Agent/API comparisons depend on the execution surfaces used here and should be read as harness-level comparisons rather than isolated causal estimates. Each configuration is evaluated with a fixed setup and single sampled run per case, so we do not estimate run-to-run variance or prompt sensitivity. \textbf{External validity.} The suite uses 21 small C++ repositories for diagnostic clarity. This supports interpretability and oracle precision, but does not cover the full diversity of languages, repository scales, or software-engineering workflows. \textbf{Conclusion validity.} Some maintainability dimensions are represented by only two or three cases. Accordingly, dimension-level differences are best interpreted as suite-level diagnostic trends rather than high-power statistical estimates.


\section{Related Work}

\vspace{.3em}
\noindent{\bf Software Engineering Benchmarks for LLMs.}
A large body of work evaluates coding systems primarily through behavioral correctness. Early benchmarks such as HumanEval~\cite{chen2021codex}, MBPP~\cite{austin2021mbpp}, and CodeContests~\cite{li2022codecontests} focus on synthesis tasks whose outputs are judged by hidden tests. More recent benchmarks broaden the setting from isolated functions to realistic codebase tasks while retaining correctness as the main signal. RepoBench~\cite{liu2023repobench} introduces repository-level context for cross-file completion, LiveCodeBench~\cite{jain2024livecodebench} emphasizes contamination-free and continually refreshed evaluation, SWE-bench~\cite{jimenez2024swebench} measures issue resolution in real GitHub repositories, and EvoCodeBench~\cite{li2024evocodebench} evaluates repository-grounded code generation under evolving benchmark updates.

This line of work has since expanded along several axes. SWE-bench Verified~\cite{chowdhury2024swebenchverified}, SWE-bench Pro~\cite{deng2025swebenchproaiagents}, and Multi-SWE-bench~\cite{zan2025multiswebench} strengthen issue-resolution evaluation through improved quality control, difficulty, and multilingual coverage. SWE-Evo~\cite{thai2025sweevo} and Commit-0~\cite{Zhao2024Commit0} move toward longer-horizon development workflows, while NL2Repo~\cite{ding2026nl2repobenchlonghorizonrepositorygeneration} studies full repository generation from natural-language specifications. Automated pipelines such as SWE-rebench~\cite{badertdinov2025swerebench} and SWE-bench Live~\cite{zhang2025swebenchgoeslive} further improve realism. .

These works are indispensable for measuring task-completion capability, but their primary signal remains test behavior. \textsc{NITR} instead evaluates whether a behaviorally correct repository edit preserves maintainability. It relies on expert-curated probes to ensure data quality, so that each case cleanly encodes a specific software engineering principle and maintainability pressure.


\vspace{.3em}
\noindent{\bf Long-Horizon Software Evolution Benchmarks.}
A second line of work pushes evaluation beyond one-shot issue resolution toward broader software-development behavior and continuous repository change. DevBench~\cite{li2024devbench} evaluates multiple stages of the software-development lifecycle, SWE-EVO~\cite{thai2025sweevo} and SWE-CI~\cite{chen2026sweci} study multi-step repository evolution and continuous-integration-style maintenance, SlopCodeBench~\cite{slopcodebench} examines code erosion as agents iterate, and EvoClaw~\cite{deng2026evoclawevaluatingaiagents} evaluates continuous software evolution through executable milestone sequences that stress sustained system integrity over time. 

These efforts primarily evaluate agents under continuous task evolution measured by test behaviors, whereas \textsc{NITR} evaluates whether a behaviorally correct repository edit preserves maintainable structure in a way that can be diagnosed precisely.

\vspace{.3em}
\noindent{\bf LLMs for Design Quality and Maintainability Analysis.}
Maintainability has long been studied in software engineering through code and design quality perspectives, including coupling and cohesion metrics~\cite{chidamber1994metrics}, design principles such as SOLID~\cite{martin2003agile}, and technical debt as a way to reason about future maintenance cost~\cite{cunningham1992wycash}. 

Recent software engineering work has begun to study LLMs on maintainability and quality-oriented tasks beyond functional correctness~\cite{sun2025qualityassurance,molison2025maintainability}. Recent examples include LLM-based localization of code design issues~\cite{llmdetectdesignissues}, prompting-based detection of SOLID principle violations~\cite{pehlivan2025solidyetempiricalstudy}, evaluation of LLMs for fixing maintainability issues in real-world projects~\cite{nunes2025evaluatingeffectivenessllmsfixing}, context-enriched quality-aware code review generation~\cite{laura}, and industrial experience with LLM-based automated code review~\cite{Ramesh2025AutomatedCR}.

These efforts study how LLMs can \emph{analyze}, \emph{review}, or \emph{repair} maintainability-related problems. {\sc NITR} differs from them by evaluating whether maintainability is preserved during code generation itself. Its distinguishing mechanism is the use of hidden structural oracles tied to explicit repository-evolution pressures.

\section{Conclusion}

\textsc{NITR} is ultimately a step toward a broader shift in how we evaluate LLM for software engineering: from asking whether models can produce correct code, to asking whether they can participate responsibly in the long-term evolution of real repositories. 
As coding agents move from isolated completions to sustained collaboration with developers, the central question is no longer only correctness, but whether generated changes preserve the structural conditions that make future change possible. 
We hope \textsc{NITR} helps make maintainability a first-class objective in both evaluation and system design, and encourages the next generation of coding agents to be judged not only by what they can build today, but by whether what they build remains worth extending tomorrow.

\bibliographystyle{ACM-Reference-Format}
\bibliography{reference}

\end{document}